\documentclass[11pt,a4paper]{scrartcl}
\pdfoutput=1

\usepackage[utf8]{inputenc}
\usepackage[T1]{fontenc}
\usepackage{lmodern}

\usepackage{graphicx}
\usepackage{epsfig}
\usepackage{float}
\usepackage{cite}
\usepackage{amssymb}
\usepackage{amsmath}
\usepackage{url}
\usepackage{color}
\usepackage{cancel}
\usepackage{nicefrac}
\usepackage[normalem]{ulem}

\usepackage[format=plain,labelfont={bf},font={small}]{caption}

\usepackage{hyperref}

\setkomafont{disposition}{\normalcolor\normalfont\bfseries}

\newcommand{\alphas}{\alpha_{\mathrm{s}}}

\newcommand{\reffig}[1]{figure~\ref{#1}}
\newcommand{\reffigs}[1]{figures~\ref{#1}}
\newcommand{\Reffig}[1]{Figure~\ref{#1}}
\newcommand{\Reffigs}[1]{Figures~\ref{#1}}

\newcommand{\POWHEGBOX}{\texttt{POWHEG-BOX}}

\begin{document}

\renewcommand*{\thefootnote}{\fnsymbol{footnote}}

\begin{center}
	{\Large \textbf{Dark matter pair production in the MSSM and in simplified dark matter models at the LHC}}\\
	\vspace{.7cm}
	Christoph Borschensky\footnote{\texttt{christoph.borschensky@uni-tuebingen.de}},
	Gabriele Coniglio\footnote{\texttt{gabriele.coniglio@uni-tuebingen.de}},
	Barbara J\"ager\footnote{\texttt{jaeger@itp.uni-tuebingen.de}}

	\vspace{.3cm}
	\textit{
		Institute for Theoretical Physics, University of T\"ubingen, Auf der Morgenstelle 14, 72076 T\"ubingen, Germany\\
	}
\end{center}   

\renewcommand*{\thefootnote}{\arabic{footnote}}
\setcounter{footnote}{0}

\vspace*{0.1cm}
\begin{abstract}
	We study the quantitative features of dark matter pair production at the LHC in simplified dark matter models and in the MSSM taking next-to-leading order QCD corrections and parton-shower effects fully into account. Among the large space of dark matter models, we focus on two particular models where a fermionic dark matter candidate interacts with the Standard Model via the exchange of either a vector mediator in the $s$-channel or coloured scalar mediators in the $t$-channel. We find that the simplified models are capable of reproducing the predictions of the MSSM to some extent in simplified supersymmetric scenarios, but lack the complexity to describe the complete theory over the full supersymmetric parameter space.
\end{abstract}

\section{Introduction}\label{s:intro}
The idea that dark matter (DM) is the main component of the matter content of the universe has been around for several decades and is well supported by astrophysical observations at large scales \cite{Bertone:2004pz}. These observations allowed to understand the abundance and distribution of DM in the universe and in many galaxies. Nevertheless, the composition of DM is basically unknown, as all attempts to detect it -- directly or indirectly -- have so far been unsuccessful \cite{Undagoitia:2015gya, Gaskins:2016cha}. While there could be some contributions to DM from the Standard Model (SM) of particle physics in the form of neutrinos as hot dark matter or from large astrophysical objects consisting of protons and neutrons, it is generally assumed that these are only subleading effects and the dominant contribution to DM is a non-baryonic type of matter which is composed of a new type of particles as predicted within many models of physics beyond the SM.

Several particles have been proposed as DM candidates. In the framework of non-relativistic or cold DM, which shows the best agreement with observations \cite{Adam:2015rua}, an important class of candidate particles are the so-called \emph{weakly interacting massive particles (WIMPs)} (for a recent review of the WIMP paradigm, see e.g.\ \cite{Arcadi:2017kky}). The interest in this kind of particles has been enhanced by the so called ``WIMP miracle'', a remarkable coincidence between a simple estimate for the relic density of DM in the universe and its measured value:  If one assumes the existence of a dark matter particle with a mass of the order of 100~GeV that interacts with the SM via an annihilation process mediated by the electroweak force, one finds that the calculated value of the relic density is close to the measured one of $\Omega_{\mathrm{DM}}h^2 = 0.1186 \pm 0.0020$ \cite{Tanabashi:2018oca,Ade:2015xua}, which corresponds to roughly one quarter of the total matter-energy content of the universe.

The various models proposed in the WIMP framework can be very different in terms of their complexity. Some complete theories formulated as extensions of the SM to solve several of its problems provide DM candidates with very specific properties, for example supersymmetry (SUSY)~\cite{Jungman:1995df} or theories predicting the existence of axions \cite{Ipser1983}. Effective field theories (EFTs) extend the SM by higher-dimensional operators in a model-independent way and can be formulated specifically to address the DM problem, aiming to locate the energy scale and the main characteristics of DM collider phenomenology. Simplified models can be considered middle ground between these approaches, as they are built having more specific types of particles and interactions in mind than in EFTs but do not refer to the complicated parameter space of a complete theory. Each of these approaches has its advantages and drawbacks: Complete theories present an unwieldily large parameter space, while simpler models may not be able to reproduce the complete DM and non-DM phenomenology encountered in the complex environment of a hadron collider and may lead to theoretical inconsistencies such as non-renormalisability and unitarity violation~\cite{Englert:2016joy}. 

A UV-complete theory that includes a DM candidate is the Minimal Supersymmetric Standard Model (MSSM). In many MSSM scenarios, the lightest neutralino, being a mixed state of the superpartners of the neutral Higgs bosons and the neutral gauge bosons of the electroweak interaction, is the lightest supersymmetric particle (LSP). With the assumption of $R$-parity conservation, it becomes an ideal DM candidate, because it is massive, interacts weakly, and is stable. The latter property is also important  from the point of view of collider physics, as the decay chains of all heavier SUSY particles which are produced in a collision result in the LSP (possibly accompanied by additional SM particles), effectively producing DM whenever SUSY particles are created and therefore opening up the possibility to study it thoroughly at colliders, should SUSY be realised in nature.
The main disadvantage of the MSSM is its large parameter space which in the most general case entails more than hundred free parameters. For this reason, it may be useful to consider simplified models that are based on only the particles and parameters strictly necessary to investigate the DM problem. A prominent example is provided by considering a Dirac fermion as DM particle that interacts with SM particles via mediator particles. A vector mediator results in an $s$-channel topology, while a set of coloured scalar mediators gives rise to $t$-channel topologies. According to these topologies, the simplified models are referred to as $s$- and $t$-channel models, respectively, and they each resemble production modes for neutralino pairs in the framework of the MSSM. 

In addition to astrophysical searches, the experimental collaborations at the Large Hadron Collider (LHC) are aiming to detect signatures for the production of DM in high-energy collisions of protons, the most prominent of them being mono-$X$ searches for large amounts of missing transverse energy, $E_{T}^{\mathrm{miss}}$, accompanied by hard recoiling particles like jets or vector bosons. A review of these searches, conducted by the CMS and the ATLAS experiments, can be found in \cite{Kahlhoefer:2017dnp}.

In this work, we consider two simplified models for DM production at the LHC, aiming to see if they exhibit similarities to more complex theories. Our DM candidate is a fermion, interacting with the SM via a vector mediator or via massive scalars. The DM production processes in our models are similar to those producing neutralinos in the MSSM, the UV-complete theory that we consider as reference for our comparison.

Our paper is structured as follows: in section~\ref{s:theoryover} we present two simplified models for DM pair production at the LHC, highlight relevant points of DM production in the MSSM, and discuss the parameter space of the models we consider. Successively, in section~\ref{s:numerics} we describe the implementation of the simplified models in the framework of the \POWHEGBOX{} and present a numerical analysis of these models at next-to-leading order (NLO) in the strong coupling matched to parton showers, where we compare these results with those predicted by restricted supersymmetric models. We conclude in section~\ref{s:conclusion}.
%

\section{Theory overview}\label{s:theoryover}
Let us first introduce the Lagrangians of the models that we employ for a description of DM. These Lagrangians can be used to derive Feynman rules for the interactions between the SM and the dark sector and to eventually predict the rate of DM particles being produced at the LHC. After sketching the simplified models, we briefly recap those features of the MSSM that are relevant for DM production.

\subsection{Simplified dark matter models}\label{s:theoryover:sub1}
Simplified dark matter models offer a way to study the phenomenology of DM at hadron colliders with a relatively small number of parameters. Their applicability is, however, limited to cases where only the DM particle itself and one or a few mediators between the dark and the SM sector enter the considered reaction, while additional heavy particles as predicted by many UV-complete theories beyond the SM are out of reach of the LHC and assumed not to interfere significantly with the DM dynamics. 
As in the context of SUSY models the expression ``simplified model'' is sometimes used to refer to simplified scenarios of the full SUSY model where only a subset of the full parameters and fields is considered, we want to clarify that in this work the term ``simplified models'' solely refers to genuine simplified dark matter models. 

We focus on two simplified models which differ in the way the SM particles couple to DM, referred to as $s$-channel and $t$-channel models, and assume that the DM field, in the following denoted by $\chi$, is a fermion. Unless stated otherwise,  $\chi$ is assumed to be a Dirac fermion, as in many experimental analyses. In addition we also discuss the Majorana case in our numerical analysis in section~\ref{s:numerics:sub4:subsub1} to establish a link to the MSSM where the neutralino represents a Majorana fermion.  
We furthermore limit our analysis to the case where the only SM particles the mediators couple to are quarks, while couplings to leptons are disregarded for this study, since we are only interested in the strong production of missing energy signals. The DM field is assumed to be a singlet under the SU(3)$_C\times$SU(2)$_L\times$U(1)$_Y$ gauge group of the SM, so at least one additional mediating field leading to an interaction between the SM and the dark sector is required.

\subsubsection{Models with an $s$-channel mediator}\label{s:theoryover:sub1:subsub1}
The simplest way to add an interaction between SM and DM fields is via the $s$-channel exchange of a massive scalar, pseudoscalar, or vector mediator field which itself is also a singlet under the SM gauge group. For instance, the vector mediator could originate from a model where the SM is extended by a new U(1)' gauge symmetry which is spontaneously broken in order for the mediator to become massive. 
For scalar and pseudoscalar mediators the interaction terms in the Lagrangian take the general form~\cite{Cotta2014, Abdullah2014} 
\begin{align}
	\mathcal{L}_S &= -g_\chi^S\bar{\chi}\chi S - \sum_q g_q^S \bar{q}q S,\\
	\mathcal{L}_P &= -i g_\chi^P\bar{\chi}\gamma_5\chi P - \sum_q g_q^P \bar{q}\gamma_5 q P,
\end{align}
respectively, with $S$ the scalar mediator field, $P$ the pseudoscalar mediator field, $q$ the quark fields with $q = u,d,c,s,t,b$, and $g_i^{S(P)}$ (with $i = q,\chi$) denoting scalar (pseudoscalar) Yukawa coupling constants. 
The interaction term in the Lagrangian for a vector mediator~\cite{Dudas2009} is given by 
\begin{align}
	\mathcal{L}_{V} &= \bar{\chi}\gamma^\mu\left[g_\chi^V-g_\chi^A\gamma_5\right]\chi V_\mu + \sum_q \bar{q}\gamma^\mu\left[g_q^V-g_q^A\gamma_5\right]q V_\mu, 
\end{align}
with $V$ the vector mediator field, and $g_i^{V(A)}$ (with $i = q,\chi$) denoting vector (axial-vector) coupling constants.

We note that minimal flavour violation (MFV) \cite{Chivukula:1987py,Hall:1990ac,Buras:2000dm,DAmbrosio:2002vsn} dictates that the Yukawa couplings for quarks $g_q^{S(P)}$ be proportional to the masses of the quarks multiplied by a flavour-independent factor which can differ for the up- and down-type flavours, see e.g.\ the comments related to MFV of \cite{Abdallah:2015ter}. Because of the smallness of the light quark masses and, thus, the relevant Yukawa couplings, DM production processes involving (pseudo-)scalar mediators in the $s$-channel at tree level are strongly suppressed. The main contribution of such mediators to DM production at hadron colliders rather stems from gluon fusion via loops of heavy quarks \cite{Haisch:2012kf,Fox:2012ru,Haisch:2013ata}, which will not be discussed in this work. In the following we focus entirely on an $s$-channel model with a vector mediator of mass $M_V$. 

\begin{figure}[tp]
	\centering
	\begin{tabular}{ccc}
		\includegraphics[width=0.3\textwidth]{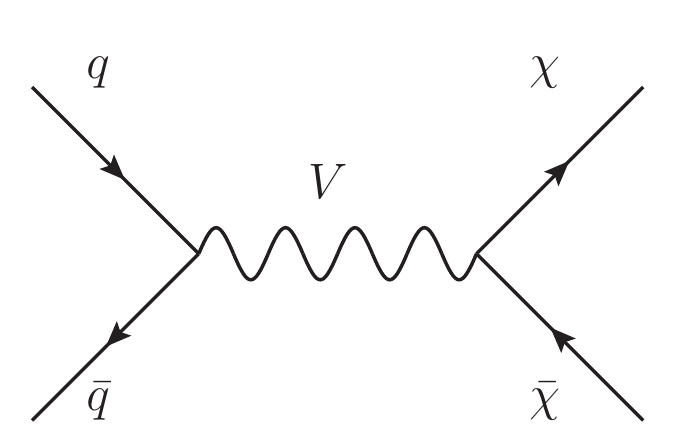} & \includegraphics[width=0.3\textwidth]{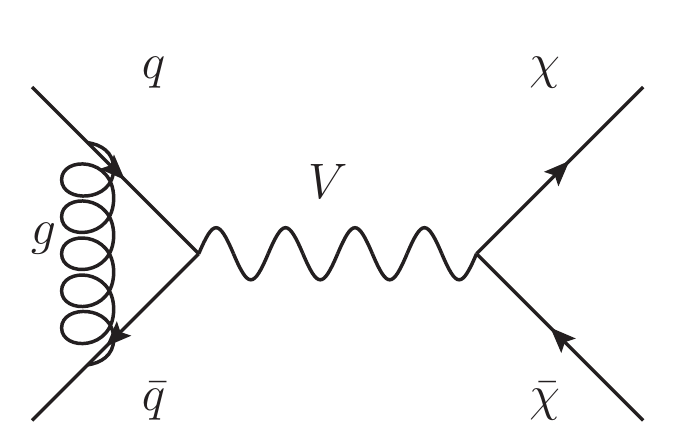} & \includegraphics[width=0.3\textwidth]{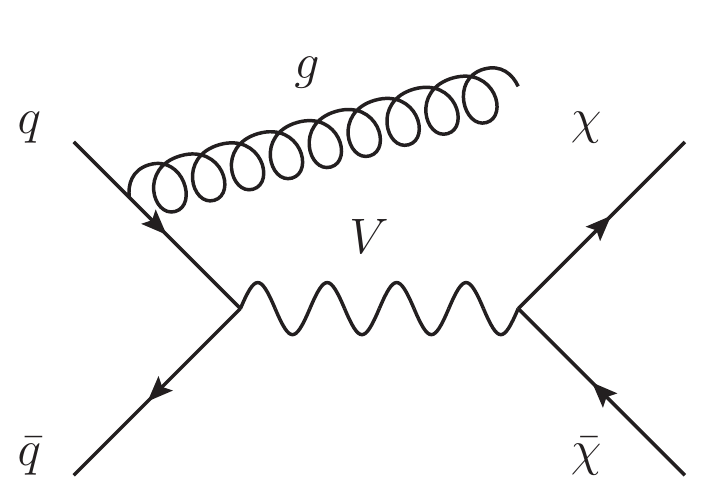}\\
		(a) & (b) & (c)
	\end{tabular}
	\caption{Feynman diagrams for DM pair production in the $s$-channel model at LO (a) and representative diagrams for the virtual (b) and real-emission (c) corrections.}
	\label{schandiags}
\end{figure}

Next-to-leading order QCD corrections to inclusive DM pair-production at the LHC in the framework of such an $s$-channel model, $pp \to \chi\bar{\chi} + X$, are of a relatively simple structure, as the only colour-charged particles in this class of reaction are the initial-state (anti-)quarks. Hence, the cross section is very similar to the well-known Drell-Yan process as illustrated by \reffig{schandiags}. Various calculations for the NLO-QCD corrections to DM pair production in the framework of a simplified $s$-channel model exist, see, for instance, Ref.~\cite{Backovic:2015soa}. In order to have full flexibility we re-computed these corrections and implemented them in a code of our own that is based on the \POWHEGBOX{} formalism. The technical details of our respective calculation are discussed in section~\ref{s:numerics:sub2}.

\subsubsection{Models with $t$-channel mediators}\label{s:theoryover:sub1:subsub2}
The case of the $t$-channel exchange of a mediator between the SM and the DM fields of a scattering process is more involved, as it requires the mediator to be in the same representation of the SM gauge group as the initial-state quarks. We discuss the case of  scalar mediators carrying colour and flavour. The interaction Lagrangian including the SM and DM fields in such a model respecting the full gauge symmetry of the SM and assuming that all Yukawa couplings are flavour diagonal is of the form \cite{An:2013xka,Abercrombie:2015wmb,Ko:2016zxg}
\begin{align}
	\mathcal{L}_{\tilde{Q}} &= -\left[\lambda_{Q_L} \bar{\chi}\left(\tilde{Q}^\dag_L \cdot Q_L\right) + \lambda_{u_R}\tilde{Q}^*_{u_R}\bar{\chi} u_R + \lambda_{d_R}\tilde{Q}^*_{d_R}\bar{\chi} d_R+\mathrm{h.c.}\right]\notag\\
	&= -\left[\lambda_{Q_L}\left(\tilde{Q}^*_{u_L}\bar{\chi} u_L+\tilde{Q}^*_{d_L}\bar{\chi} d_L\right) + \lambda_{u_R}\tilde{Q}^*_{u_R}\bar{\chi} u_R + \lambda_{d_R}\tilde{Q}^*_{d_R}\bar{\chi} d_R+\mathrm{h.c.}\right].
\end{align}
Here $\tilde{Q}_L = (\tilde{Q}_{u_L}, \tilde{Q}_{d_L})^T$ and $Q_L = (u_L, d_L)^T$ are SU(2)$_L\times$U(1)$_Y$ doublets, with $\tilde{Q}_{u_L}$, $\tilde{Q}_{d_L}$, $\tilde{Q}_{u_R}$, $\tilde{Q}_{d_R}$ the scalar mediator fields, $u_{L/R}$, $d_{L/R}$ the left- and right-handed up- and down-type quark fields. Flavour indices are suppressed and we generically  write ``$u$''$ = u,c,t$ and ``$d$''$ = d,s,b$, and $\lambda_{Q_L}$, $\lambda_{u_R}$, while $\lambda_{d_R}$ are the Yukawa couplings to the left- and right-handed quark fields. Due to the SU(2)$_L\times$U(1)$_Y$ symmetry of the mediators, the left-handed quark coupling $\lambda_{Q_L}$ is identical for up- and down-type flavours. In the most general case, each mediator field has a different mass for each flavour and ``chirality'' state, $M_{\tilde{Q}_L}$, $M_{\tilde{Q}_{u_R}}$, and $M_{\tilde{Q}_{d_R}}$. We will subsequently refer to this model as the ``$t$-channel model''. It should be noted that while in the $s$-channel model the mediator can be lighter than the DM particle, the $t$-channel mediators must be heavier than the DM particles to render them stable.

\begin{figure}[tp]
	\centering
	\begin{tabular}{ccc}
		\includegraphics[width=0.3\textwidth]{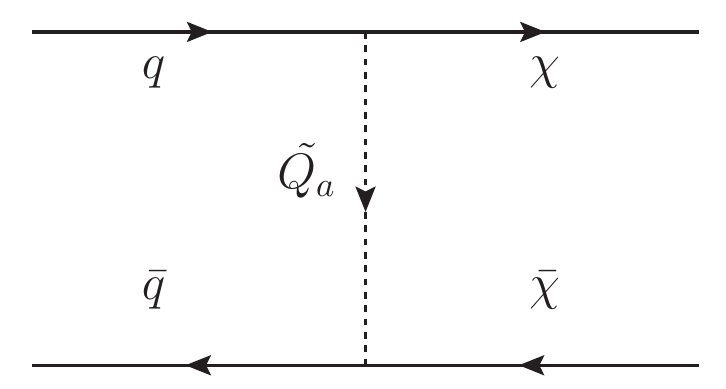} & \includegraphics[width=0.3\textwidth]{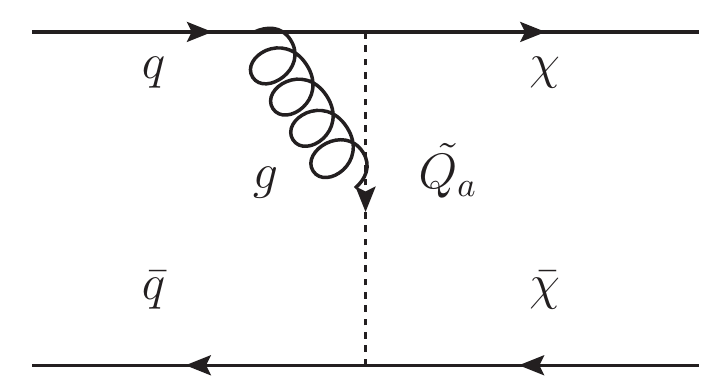} & \includegraphics[width=0.3\textwidth]{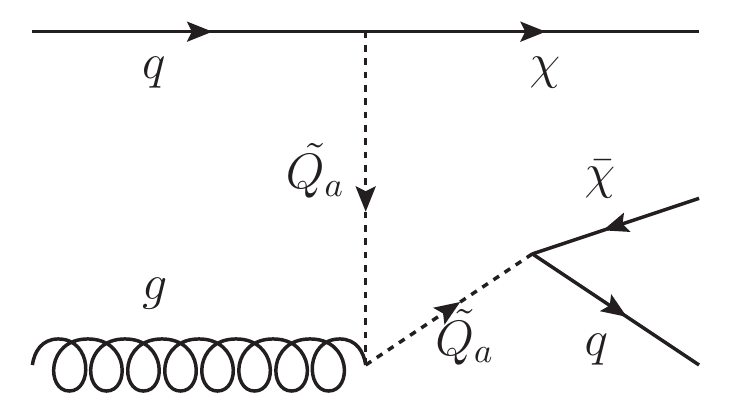}\\
		(a) & (b) & (c)
	\end{tabular}
	\caption{Feynman diagrams for DM pair production in the $t$-channel model at LO order (a) and a representative diagram for the virtual corrections (b). In (c), a representative diagram for the real-emission corrections is shown where a scalar mediator can become resonant.}
	\label{tchandiags}
\end{figure}

In analogy to the $s$-channel model, NLO-QCD corrections to the process $pp \to \chi\bar{\chi} + X$ can be calculated. While the computation of the virtual corrections is straightforward -- with the major difference to the $s$-channel case being that more loop diagrams are involved due to the colour charge of the mediator --, the real corrections require special consideration because of the appearance of on-shell resonances as sketched in \reffig{tchandiags}~(c). If the scalar particle $\tilde{Q}_a$ is on-shell, the real-emission process $q\bar q \to \chi\bar{\chi} q$ can be seen as a tree-level production of $\chi \, \tilde{Q}_a $ followed by the decay of $\tilde{Q}_a$ into $q \, \bar{\chi}$ and not as a real-emission correction to the process of interest, i.e.\ $q\bar q \to \chi\bar{\chi}$. For this reason, the on-shell contributions have to be subtracted in order to avoid double-counting and to preserve the validity of the perturbative expansion. A more detailed discussion of the treatment of such on-shell contributions will be given in section~\ref{s:numerics:sub2:subsub1}.

\subsection{The MSSM}\label{s:theoryover:sub2}
As mentioned above, in the MSSM with $R$-parity conservation the LSP constitutes a suitable DM particle candidate.  In many SUSY scenarios, the LSP is the lightest neutralino which is a Majorana fermion that is only weakly charged. In this work, we thus limit our discussion to the case where the DM candidate of SUSY is the lightest neutralino. The DM pair production process in the MSSM then corresponds to the production of neutralino pairs, whose NLO-QCD corrections, including also SUSY-QCD contributions, for the production at hadron colliders were first calculated and presented in \cite{Beenakker:1999xh}.
We note that the phenomenology of DM pair production in the context of the MSSM is significantly influenced by the structure of the neutralino mixing matrix $N_{ij}$  ($i,j=1,\ldots 4$) that determines how the four neutralinos of the MSSM, $\tilde{\chi}_i^0$ ($i=1,\ldots 4$), are obtained from the bino, wino, and higgsino fields.

\begin{figure}[tp]
	\centering
	\begin{tabular}{ccc}
		\includegraphics[width=0.3\textwidth]{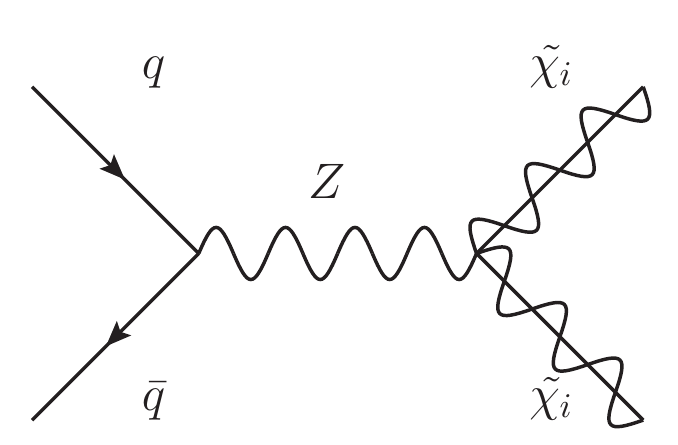} & \includegraphics[width=0.3\textwidth]{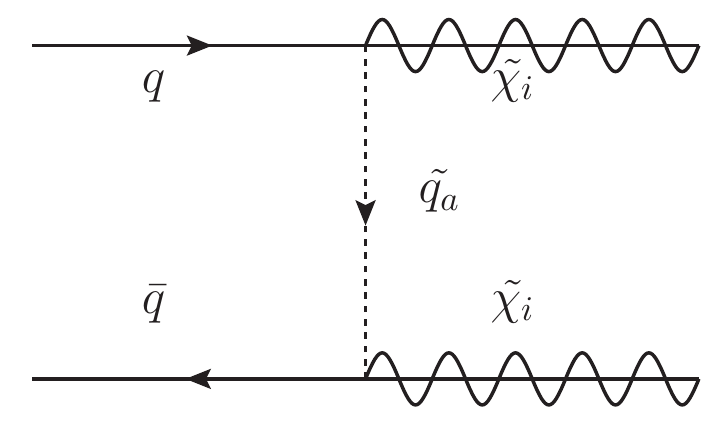} & \includegraphics[width=0.3\textwidth]{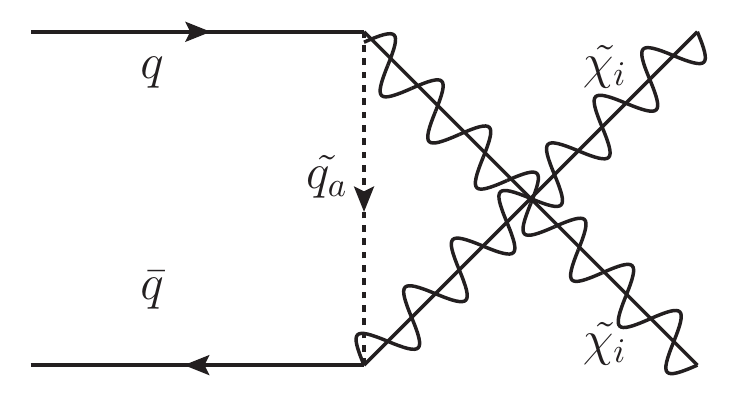}\\
		(a) & (b) & (c)
	\end{tabular}
	\caption{Feynman diagrams for the production of a neutralino pair in the MSSM. Since the neutralino is a Majorana fermion,  in addition to the $s$- (a) and $t$-channel (b) diagrams, a $u$-channel diagram (c) is also included. } 
	\label{mssmdiags}
\end{figure}

A main goal of this work then is to compare the DM simplified models introduced above to the case of neutralino pair production in the MSSM at the LHC. \Reffig{mssmdiags} shows representative tree-level Feynman diagrams for neutralino-pair production in the MSSM that indeed comprise topologies similar to those encountered for the simplified models illustrated in \reffigs{schandiags} and \ref{tchandiags}.

As briefly mentioned above, a major drawback of the MSSM is its large and unconstrained parameter space of more than hundred free parameters. For this reason, one often refers to simpler SUSY scenarios, such as the constrained MSSM (CMSSM) \cite[and references therein]{Ellis2016} or the phenomenological MSSM with ten parameters (pMSSM10) \cite[and references therein]{deVries:2015hva}, where only a handful of parameters, defined at a specific energy scale, is used as an input to generate the entire SUSY spectrum using renormalisation group methods. 

The input parameters of the CMSSM are the universal SUSY-breaking mass parameters for scalar and fermionic superpartners, $m_0$ and $m_{\nicefrac{1}{2}}$, respectively, the trilinear scalar coupling parameter $A_0$, the ratio of the Higgs vacuum expectation values of the two Higgs doublets $\tan\beta$, and the sign of the Higgs coupling parameter $\mu$. The value of each of these parameters is defined at the grand unification scale. 

In the pMSSM10 the parameters are chosen to be the squark masses for the three generations, $m_{\tilde{q}_1}$, $m_{\tilde{q}_2}$ and $m_{\tilde{q}_3}$, where the first two are taken to be equal; the gaugino masses $M_1$, $M_2$ and $M_3$; a slepton mass term $m_{\tilde{l}}$; the trilinear mixing parameter $A_b$ and $A_t$ for the bottom and the top squarks, the  higgsino mass parameter $\mu$, the pseudo-scalar mass $M_A$ and the ratio of the Higgs vacuum expectation values $\tan \beta $. Contrary to the CMSSM, all the pMSSM10 parameters are defined at the scale $M_{\mathrm{SUSY}} = \sqrt{m_{\tilde{t}_1} m_{\tilde{t}_2}}$, where $ m_{\tilde{t}_1}$ and $m_{\tilde{t}_2}$ are the masses of the stop eigenstates.

\subsection{Parameters of the simplified models}\label{s:theoryover:sub3}
In the MSSM we always consider the lightest neutralino $\tilde{\chi}^0_1$ to be the DM candidate. For comparisons of  simplified models with the MSSM, we therefore identify the mass $m_\chi$ of the respective DM particle $\chi$  with the mass $m_{\tilde{\chi}^0_1}$ of the lightest neutralino $\tilde{\chi}^0_1$ of the MSSM, 
\begin{equation}
	m_\chi = m_{\tilde{\chi}^0_1}.
\end{equation}
In the $t$-channel model, we additionally assume that the mediators of all flavours are degenerate in their mass ($M_{\tilde{Q}_L} = M_{\tilde{Q}_{u_R}} = M_{\tilde{Q}_{d_R}} \equiv M_{\tilde{Q}}$), and set $M_{\tilde{Q}}$ to the average of the ten light-flavoured squark masses,
\begin{equation}
	M_{\tilde{Q}} = \frac{m_{\tilde{u}_L}+m_{\tilde{u}_R}+m_{\tilde{d}_L}+m_{\tilde{d}_R}+m_{\tilde{c}_L}+m_{\tilde{c}_R}+m_{\tilde{s}_L}+m_{\tilde{s}_R}+m_{\tilde{b}_1}+m_{\tilde{b}_2}}{10}.
\end{equation}
For the vector mediators of the $s$-channel model, we consider two different masses ($M_V = 1$~TeV and $M_V = 10$~TeV) to cover the phenomenologically interesting cases where (i) the mediator is too light to decay on-shell into a pair of DM particles ($M_V \lesssim 2m_\chi$), and (ii) where it is too heavy so that the on-shell effects only contribute negligibly to the process for LHC center-of-mass energies of $\sqrt{S}=13$~TeV ($2m_\chi \ll M_V \sim \sqrt{S}$).

\begin{figure}[tp]
	\centering
	\includegraphics[width=0.5\textwidth]{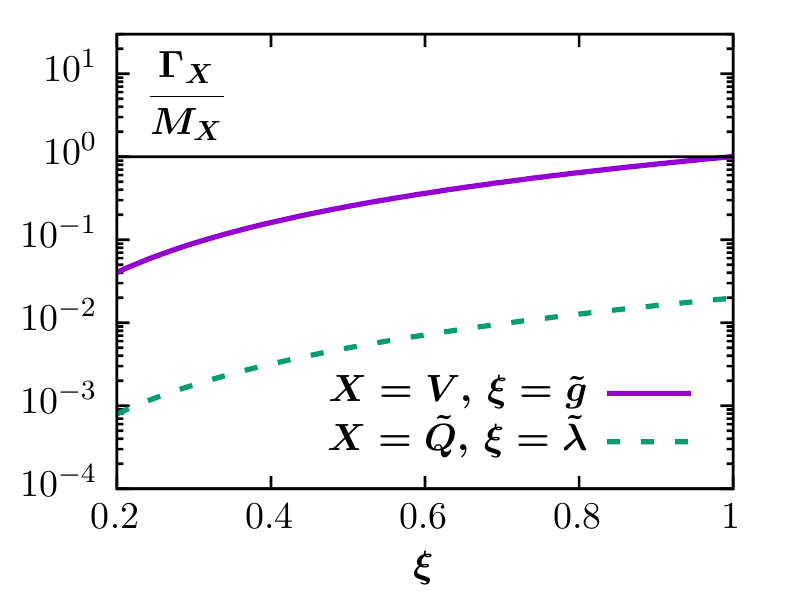}
	\caption{Ratio of mediator width $\Gamma_X$  to mediator mass $M_X$ for the two cases of a spin-1 mediator~$V$ in the $s$-channel model (solid purple) and a scalar mediator~$\tilde{Q}$ in the $t$-channel model (dashed green), as a function of the corresponding coupling ($\xi=\tilde{g}$ or $\xi=\tilde{\lambda}$). The parameters are fixed to $M_V = 10$~TeV, $M_{\tilde{Q}} = 3$~TeV, and $m_\chi = 100$~GeV. }
	\label{widthplots}
\end{figure}
In order to minimise the number of parameters of the simplified models, we choose for the $s$-channel model $g^V_{\chi/q} = g^A_{\chi/q} \equiv \tilde{g}$, and for the $t$-channel model $\lambda_{Q_L} = \lambda_{u_R} = \lambda_{d_R} \equiv\tilde{\lambda}$, irrespective of the quark flavour $q$. The couplings $\tilde{g}$ and $\tilde{\lambda}$ are chosen such that the ratio of the mediator width to its mass is not too large to justify that by rescaling the couplings the cross section changes proportionally to this rescaling. In the following we therefore set $\tilde{g} = 0.5$ and $\tilde{\lambda} = 1$. The ratio of mediator width to its mass can be calculated in the two models at leading order in the strong coupling, $\alphas$. Assuming five massless quark flavours and the above-mentioned simplifications for the couplings, for the $s$-channel model we find 
\begin{align}
	\frac{\Gamma_V}{M_V} = \frac{\tilde{g}^2}{6\pi}\Bigg\{15 &+ \Theta(M_V-2m_t)3\sqrt{1-\frac{4m_t^2}{M_V^2}}\left(1-\frac{m_t^2}{M_V^2}\right)\notag\\
	&+\Theta(M_V-2m_\chi)\sqrt{1-\frac{4m_\chi^2}{M_V^2}}\left(1-\frac{m_\chi^2}{M_V^2}\right)\Bigg\}\label{schanwidth},
\end{align}
where $\Theta(x)$ is the Heaviside step function with $\Theta(x) = 0$ for $x<0$ and $\Theta(x) = 1$ for $x\ge 0$ and $m_t = 173.0$~GeV~\cite{Tanabashi:2018oca} is the top quark mass. For the $t$-channel model we obtain 
\begin{align}
	\frac{\Gamma_{\tilde{Q}}}{M_{\tilde{Q}}} = \frac{\tilde{\lambda}^2}{16\pi}\left(1-\frac{m_\chi^2}{M_{\tilde{Q}}^2}\right)^2\label{tchanwidth}.
\end{align}
\Reffig{widthplots} illustrates the ratio of width to mass for both the mediators of the $s$- and $t$-channel models as a function of the corresponding coupling. It can be seen that with our choice of the couplings, the ratios are well below one, i.e.\ the widths are small enough for the narrow-width approximation to be valid. 
%

\section{Numerical analysis}\label{s:numerics}
In the following we present a numerical comparison of the models introduced above focusing on total cross sections and differential distributions for the process of inclusive DM pair production at the LHC,
\begin{align*}
	pp \to \chi\bar{\chi} + X, 
\end{align*}
where $\chi$ denotes either the fermionic  DM candidate of a simplified model or the lightest neutralino of the MSSM systematically accounting for radiative corrections and parton-shower effects. The comparison involves a scan over the SUSY parameter space and a suitable identification of the parameters of the simplified models with the respective MSSM parameters. Our aim is to find and study those regions of the MSSM parameter space where DM pair production can be described adequately by either of the simplified models considered in this work. We start with explaining the setup of the codes used in our analysis and then present our results.

\subsection{Generation of SUSY spectra}\label{s:numerics:sub1}
For the calculation of SUSY spectra in the CMSSM and the pMSSM10 we have used the program \texttt{SPheno~4.0.3}~\cite{Porod:2003um,Porod:2011nf} restricting ourselves to the following ranges of input parameters of the CMSSM \cite{Ajaib:2017iyl}:
\begin{gather*}
	m_0 \in [0,10]~\mathrm{TeV},~~m_{\nicefrac{1}{2}} \in [0,10]~\mathrm{TeV},~~A_0 \in [-3,3]\times m_0,~~\tan\beta\in [2,60],~~\operatorname{sign}\mu = +1,
\end{gather*}
and of the pMSSM10 \cite{deVries:2015hva}:
\begin{gather*}
	M_1 \in [-1,1]~\mathrm{TeV},~~M_2 \in [0,4]~\mathrm{TeV},~~M_3 \in [-4,4]~\mathrm{TeV},\\
	m_{\tilde{q}_1} = m_{\tilde{q}_2} \in [0,4]~\mathrm{TeV},~~m_{\tilde{q}_3} \in [0,4]~\mathrm{TeV},~~m_{\tilde{\ell}} \in [0,2]~\mathrm{TeV},\\
	M_A \in [0,4]~\mathrm{TeV},~~A_b = A_t \in [-5,5]~\mathrm{TeV},~~\mu \in [-5,5]~\mathrm{TeV},~~\tan\beta\in [1,60]. 
\end{gather*}
Additionally we required that 1) the lightest neutralino $\tilde{\chi}^0_1$ is the LSP, and 2) the mass of the lightest Higgs boson $h$ is compatible with the experimentally measured Higgs boson mass, i.e.\ it satifies $124~\mathrm{GeV} \le m_h \le 126~\mathrm{GeV}$. The DM relic abundance has not been used as a constraint for the selection of the parameter points. We, however, checked that both for the CMSSM and the pMSSM10, our parameter ranges include points with the observed relic abundance. While its numerical value has been determined very precisely by experimental measurements, its theoretical calculation is based on several assumptions \cite{Albert:2017onk}, e.g.\ only the fields available in the studied model take part in DM annihilation processes. If these assumptions are violated, e.g.\ if there are additional hidden fields, the observed relic abundance can be reached even if DM is produced over- or underabundantly in the considered model.

A total of 5000 SUSY spectra has been generated for each of the two constrained MSSM scenarios. We checked that the parameter points are sufficiently widely distributed over the respective parameter ranges. Nevertheless, some regions of the parameter space are less densely populated because of the experimental constraints we imposed.

\subsection{Code generation and numerical implementation}\label{s:numerics:sub2}
The numerical computation of cross sections and kinematical distributions, both in the MSSM and in the simplified models we consider, has been performed by means of customized codes we developed in the framework of the \POWHEGBOX{}~\cite{Nason:2004rx,Frixione:2007vw,Alioli:2010xd}. In that way we are able to provide predictions for cross sections and differential distributions at NLO-QCD accuracy, and to additionally include parton shower (PS) effects. Our predictions for neutralino pair-production are obtained with the code package for electroweakino pair-production \cite{Baglio:2016rjx}, which is publicly available from the \POWHEGBOX{} repository~\cite{pwgboxweb}. We note that $\texttt{SPheno}$ provides output for the SUSY spectra we consider in the form of  files in the SUSY Les Houches Accord (SLHA) format \cite{Skands:2003cj,Allanach:2008qq}, which can easily be interfaced with the \POWHEGBOX{} codes we are using. 

For DM pair production in simplified $s$- and $t$-channel models, we have developed a \POWHEGBOX{} implementation of our own. The main ingredients which are needed to include a process in this framework are the Born, virtual, and real-emission matrix elements as well as suitable subtraction terms for the cancelation of infrared divergencies. Both the Born and the real-emission amplitudes can be automatically generated by a tool based on \texttt{MadGraph~4}~\cite{Murayama:1992gi,Stelzer:1994ta,Alwall:2007st}. For the cancellation of infrared (soft or collinear) divergences between the virtual and real-emission amplitudes and the subtraction of initial-state collinear singularities, performed via the Frixione-Kunszt-Signer formalism~\cite{Frixione:1995ms}, the automated tool provides the spin- and colour-correlated Born amplitudes, which are necessary for the construction of the subtraction terms in the limit of soft and collinear emission of the extra parton.

The virtual one-loop amplitudes are generated by the two \texttt{Mathematica} packages \texttt{FeynArts 3.9} \cite{Hahn:2000kx} and \texttt{FormCalc 9.4} \cite{Hahn:1998yk}. The resulting numerical code is interfaced with \texttt{COLLIER 1.2} \cite{Denner:2016kdg,Denner:2002ii,Denner:2005nn,Denner:2010tr} for the numerical computation of the one-loop integrals. We renormalise external quark fields using the on-shell scheme, whereas the mediator fields and masses as well as the couplings of the $t$-channel model are renormalised in the $\overline{\mathrm{MS}}$ scheme. None of the other parameters, in particular those of the $s$-channel model, require renormalisation at NLO QCD. We have checked that for both the $s$- and $t$-channel models, all ultraviolet divergences cancel after including the proper renormalisation terms.

Throughout our calculation, we consider a hadronic centre-of-mass energy of $\sqrt{S}=13$~TeV. For the parton distribution functions (PDFs) of the initial-state protons we use the PDF4LHC15~NLO~MC set~\cite{Butterworth:2015oua}. The PDFs are accessed via the \texttt{LHAPDF 6} library~\cite{Buckley:2014ana}. We work in the five-flavour scheme, meaning that we assume five massless quark flavours to appear in the initial state and to take part in the running of $\alphas(\mu_R)$, where $\mu_R$ is the renormalisation scale. At the $Z$-boson mass $M_Z = 91.188$~GeV, the value $\alphas(M_Z) = 0.118$ is used. The strong coupling is given in the $\overline{\mathrm{MS}}$ renormalisation scheme. Both the renormalisation scale $\mu_R$ and the factorisation scale $\mu_F$ are set equal to the sum of the final-state masses of the corresponding process, i.e.\ $\mu_R = \mu_F = 2m_{\tilde{\chi}^0_1}$ or $2m_\chi$. As opposed to a dynamical scale this choice simplifies systematic comparisons between different models.

As a parton-shower generator, we use \texttt{PYTHIA~6.4.25}~\cite{Sjostrand:2006za}. We switch off all effects related to QED radiation, underlying event, and hadronisation. For the generation of jets we employ the \texttt{FastJet~3.3.0} package~\cite{Cacciari:2005hq,Cacciari:2011ma} using the anti-$k_T$ algorithm~\cite{Cacciari:2008gp} with $R = 0.4$ and $|\eta^{\mathrm{jet}}| < 4.5$.

\subsubsection{On-shell subtraction for the $t$-channel model}\label{s:numerics:sub2:subsub1}
Before discussing the results of our numerical study, we briefly want to outline the technical issue of the subtraction of on-shell resonances as required for the proper implementation of the real-emission corrections within the $t$-channel model. We closely follow \cite{Baglio:2016rjx,Gavin:2013kga,Gavin:2014yga} in that we split up the sum of our real-emission matrix elements into contributions $\mathcal{M}_r$ which can exhibit a resonance  (c.f.\ \reffig{tchandiags}~(c)) for an intermediate propagator of the $t$-channel mediator~$\tilde{Q}$,  and contributions $\mathcal{M}_{nr}$ which do not contain any resonant propagators, 
\begin{align}
	\mathcal{M}_{\mathrm{real}} = \mathcal{M}_{nr} + \mathcal{M}_{r}.
\end{align}
Would-be divergences of the resonant contributions can be regulated by a width in the respective propagators. 
In practice, we globally introduce a width in all propagators containing the $\tilde{Q}$~particle  by replacing the mass of the $t$-channel mediator with a complex expression containing the width parameter $\Gamma$, 
\begin{align}
	\frac{1}{p^2-M_{\tilde{Q}}^2} \to \frac{1}{p^2-M_{\tilde{Q}}^2+i\Gamma M_{\tilde{Q}}}.
\end{align}
We then subtract from the absolute square of the real-emission amplitude $|\mathcal{M}_{\mathrm{real}}|^2 = |\mathcal{M}_{nr}|^2 + 2\operatorname{Re}(\mathcal{M}_{nr}\mathcal{M}_r^*) + |\mathcal{M}_r|^2$ a local counterterm that is proportional to the square of the diagrams which can become resonant,  evaluated for on-shell remapped kinematics, i.e.\ where the on-shell condition $p^2 = M_{\tilde{Q}}^2$ is true. Further details on the construction of the counterterm and its implementation in a Monte-Carlo code can be found in the above-mentioned references.

The implementation of the on-shell subtraction procedure in our code has been performed by manually extracting the possibly resonant diagrams from our \texttt{MadGraph}-based routines for the real-emission amplitudes and adding the counterterm to their integrand. We then checked the stability of our subtraction procedure by varying the value of the width parameter $\Gamma$. We found that the total cross section is essentially independent of this value and therefore of the resonance, proving that our implementation is meaningful. 

\begin{figure}[tp]
	\centering
	\includegraphics[width=0.5\textwidth]{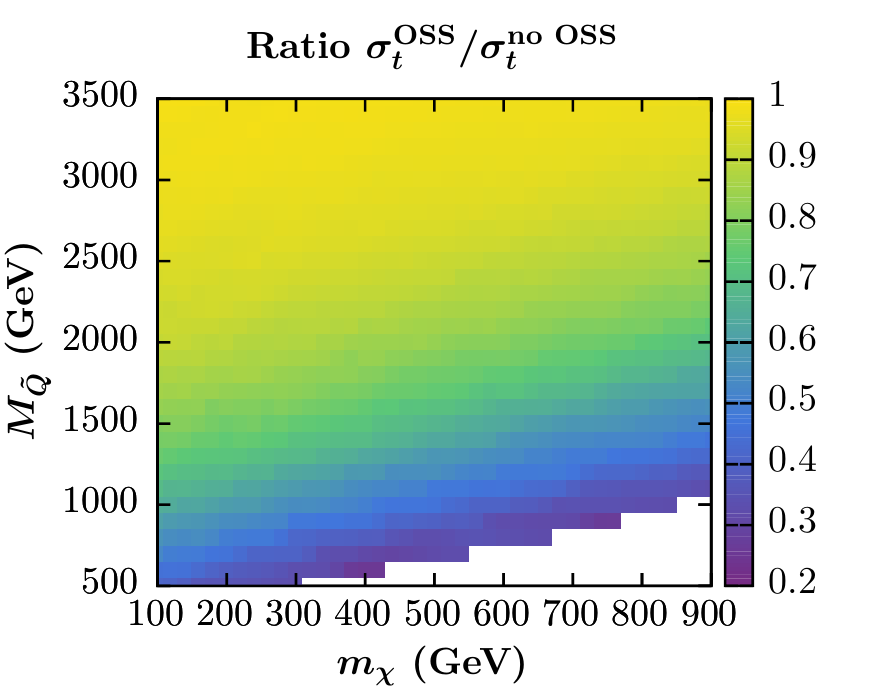}
	\caption{Impact of the subtraction of on-shell resonances in the real-emission contributions to DM pair production in the $t$-channel model. Shown is the ratio of the total cross section at NLO QCD with the subtraction of on-shell contributions ($\sigma_t^\mathrm{OSS}$) to the case ($\sigma_t^\mathrm{no~OSS}$) where the on-shell resonances are not subtracted, but regularised by a finite width only. The axes of the plot are chosen as the DM mass $m_\chi$ and the average $t$-channel mediator mass $M_{\tilde{Q}}$.}
	\label{tchanoss}
\end{figure}
In \reffig{tchanoss}, we show the effect of the on-shell subtraction on the total cross section for DM pair production in the $t$-channel model. While for the case of $M_{\tilde{Q}}\gg m_\chi$, the difference between the on-shell subtracted and the non-subtracted case is negligible, the cross sections can differ by up to a factor of five close to $M_{\tilde{Q}} \approx m_\chi$ where the on-shell contributions are the largest. For the following predictions of the $t$-channel model, all on-shell contributions have been subtracted.

\subsection{Analysis within the CMSSM}\label{s:numerics:sub3}
\begin{figure}[tp]
	\centering
	\begin{tabular}{cc}
		\includegraphics[width=0.48\textwidth]{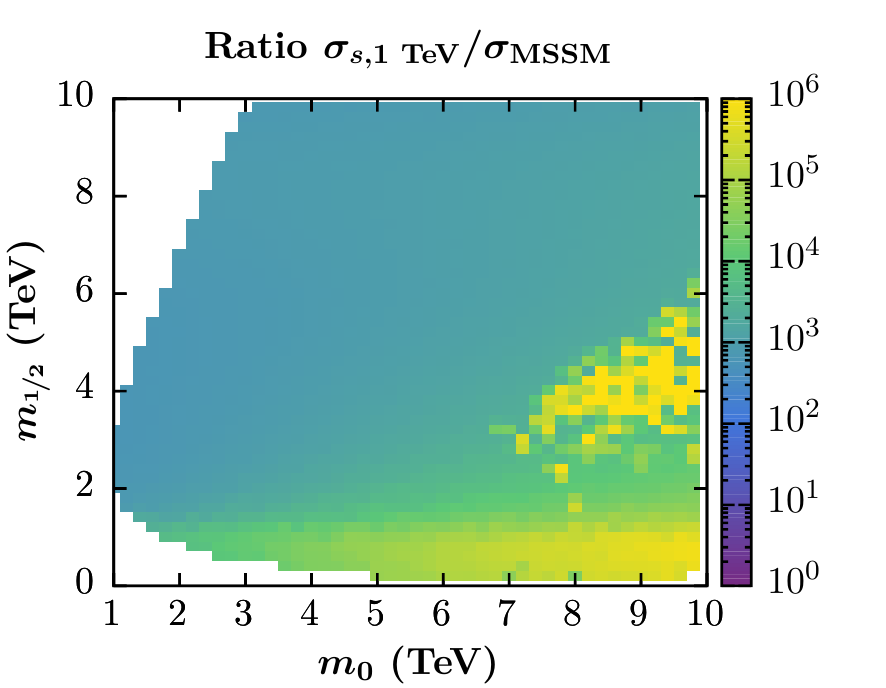} & \includegraphics[width=0.48\textwidth]{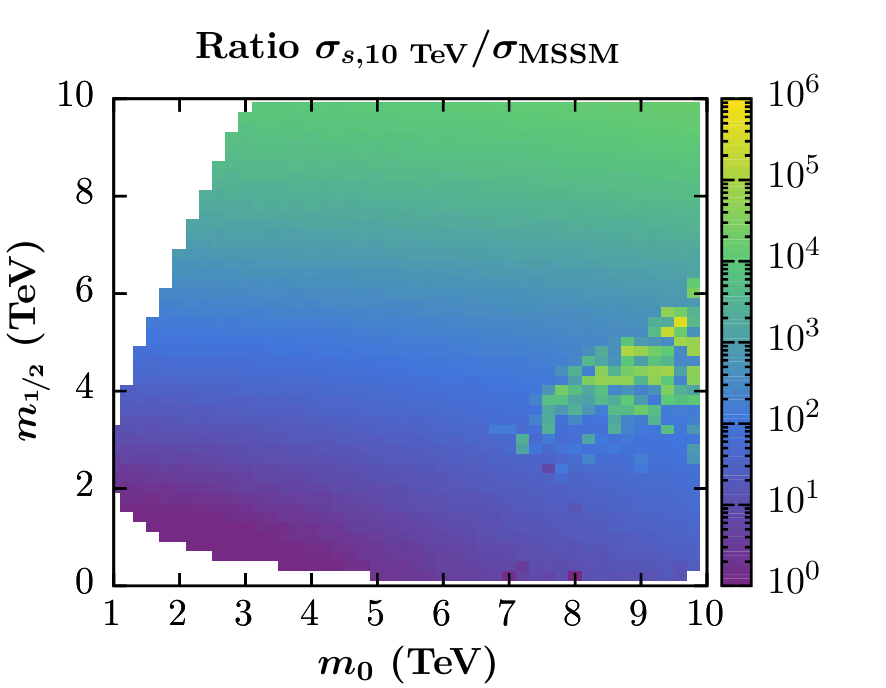}\\
		(a) & (b)\\
		\includegraphics[width=0.48\textwidth]{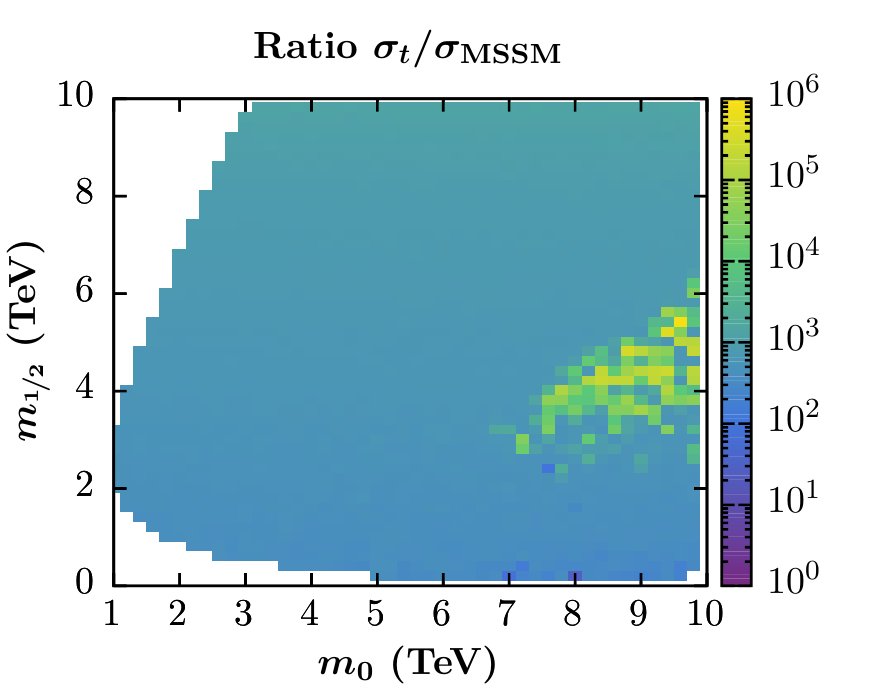} & \includegraphics[width=0.48\textwidth]{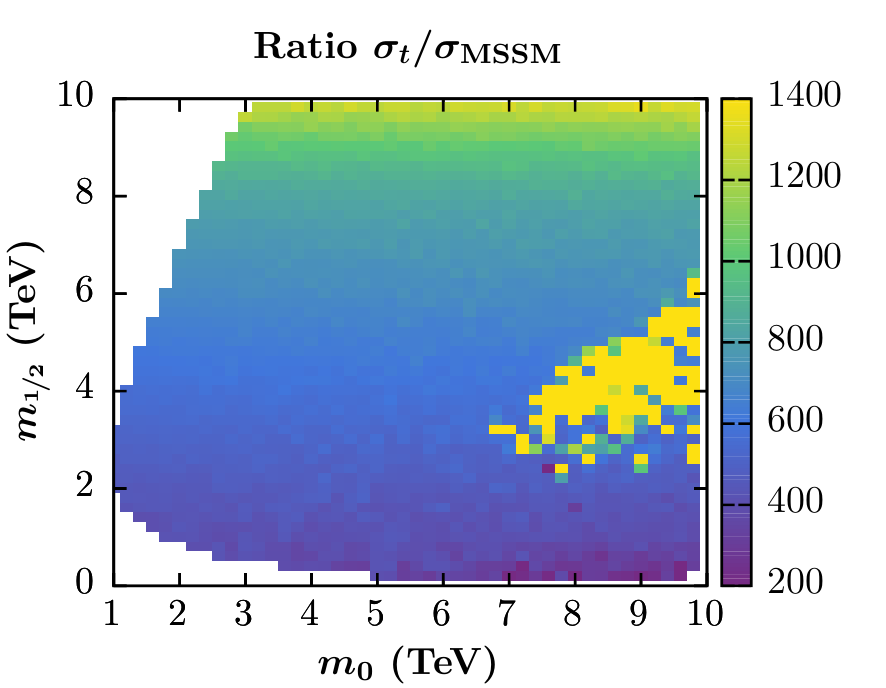}\\
		(c) & (d)
	\end{tabular}
	\caption{Ratios of total cross sections for DM pair production in the simplified models to the ones in the MSSM, in the plane of the CMSSM parameters $m_0$ and $m_{\nicefrac{1}{2}}$, with all other parameters varied implicitly. Shown are the $s$-channel models with $M_V = 1$~TeV~(a) and $M_V = 10$~TeV~(b), and the $t$-channel model with a logarithmic (c) and a linear (d) colour scale, respectively. White regions denote points excluded in the parameter space where either the LSP is charged and therefore no DM candidate (top left corner) or where radiative electroweak symmetry breaking is not possible (bottom left corner). Rough edges are an interpolation effect.}
	\label{cmssmhmplots}
\end{figure}
We start our analysis with a comparison of total cross sections in the CMSSM and the various simplified models we consider, with model parameters fixed as discussed in section~\ref{s:theoryover:sub3}. 
In \reffig{cmssmhmplots}, we show the ratios of total cross sections in the simplified models to the ones in the MSSM, with the axes being chosen as $m_0$ and $m_{\nicefrac{1}{2}}$. The other CMSSM parameters vary implicitly in the above-mentioned ranges. This results in a range for the mass of the DM candidate between approximately $100$~GeV and $5$~TeV. The magnitude of the ratios is colour-coded, with purple indicating low and yellow very high ratios. For \reffigs{cmssmhmplots}~(a), (b), and (c), the range of the ratios is chosen to be the same, as not the absolute values of the ratios are of interest for our comparison (they can be rescaled by varying the values of the couplings in the simplified models), but the relative variations of the ratios over the shown parameter range.

\Reffigs{cmssmhmplots}~(a) and (b) show the $s$-channel model with mediator masses of $M_V = 1$~TeV and $M_V = 10$~TeV, respectively. For these two cases, a clear difference in the structure of the ratios can be seen. While the case of $M_V = 1$~TeV shows a relatively flat behaviour for $m_{\nicefrac{1}{2}} > 2$~TeV with at most one order of magnitude difference between the ratios (except for a region of points with $m_0 > 7$~TeV and $2~\mathrm{TeV} < m_{\nicefrac{1}{2}} < 6~\mathrm{TeV}$ which we will discuss explicitly below), the $s$-channel model with $M_V = 10$~TeV leads to differences of up to six orders of magnitude over the shown parameter range. This difference can be explained by the resonance behaviour of the $s$-channel model: for $M_V = 2m_{\chi}$, the internal propagator of the mediator becomes resonant and thus the cross section is strongly enhanced close to this region. This can be seen in \reffig{cmssmhmplots}~(a) for low $m_{\nicefrac{1}{2}} < 2$~TeV and also in \reffig{cmssmhmplots}~(b) for high $m_{\nicefrac{1}{2}} \approx 10$~TeV. As the neutralino mass $m_{\tilde{\chi}^0_1}$ and therefore also $m_\chi$ mainly depend on $m_{\nicefrac{1}{2}}$, we find that at the lower edge of the parameter plane, $m_\chi$ is around 500~GeV, thus leading to a resonance for the $M_V = 1$~TeV case. At the upper edge, $m_\chi$ is around 5~TeV and thus causing the cross section of the simplified model, and therefore the ratio, for the $M_V = 10$~TeV case to grow significantly towards this value.

The case of the $t$-channel model is shown in \reffigs{cmssmhmplots}~(c) and (d) with a logarithmic and a linear colour scale, respectively. From these two plots it becomes clear that over the whole parameter range, the ratio varies by less than one order of magnitude, which is indicated by an almost constant colouring of the area. The linear colour scale reveals that for most of the parameter space, the ratio varies at most by a factor of three. The greater similarity to the MSSM than for the case of the $s$-channel model can be explained by the identification of the squark masses with the masses of the $t$-channel mediators. While the $s$-channel model only shares one parameter ($m_\chi$) with the MSSM, the $t$-channel model shares two ($m_\chi$, $M_{\tilde{Q}}$).

Lastly, we want to comment on the points with $m_0 > 7$~TeV and $2~\mathrm{TeV} < m_{\nicefrac{1}{2}} < 6~\mathrm{TeV}$. In this region of the parameter space, the ratio is strongly enhanced by several orders of magnitude compared to the surrounding points. Two effects play a role in this enhancement: First, $m_{\tilde{\chi}^0_1} = m_\chi$ is lower than for the surrounding points, leading to an increase of the cross sections both in the numerator and the denominator of the ratio. Second, in the parameter space that we show, there is a general suppression of the $Z$-boson contribution to the $\tilde{\chi}^0_1$ pair production process in the MSSM (\reffig{mssmdiags}~(a)). We have checked that for all of the points with the exception of the enhanced ones, $\tilde{\chi}^0_1$ essentially consists of a bino component only. This results in only the diagrams with squark exchange of \reffigs{mssmdiags}~(b) and (c) to contribute to the cross section. For the enhanced points, however, the bino component (and therefore the squark exchange diagrams) is strongly suppressed, and $\tilde{\chi}^0_1$ is mainly a mix of the two higgsino components. As the amplitude corresponding to \reffig{mssmdiags}~(a) is proportional to the difference of the absolute squared values of the neutralino mixing matrix elements for the higgsino components \cite{Rosiek:1989rs}, i.e.\ $\mathcal{M}_{Z\text{-boson}} \propto |N_{14}|^2-|N_{13}|^2$, the approximately equal size of $|N_{13}|$ and $|N_{14}|$ leads to a suppression of the $Z$-boson diagram also for these parameter points. Subsequently, due to a suppression of all diagrams of \reffig{mssmdiags} in this parameter range, the MSSM cross section in the denominator of the ratio becomes very small, leading to the strong enhancement of the ratio itself. In the following, we will refer to this effect as the ``higgsino mixing-matrix suppression''. We note that this mixing-matrix suppression leading to a decrease of the MSSM cross sections is purely a SUSY effect and does not exist for the simplified model cross sections.

\subsection{Analysis within the pMSSM10}\label{s:numerics:sub4}
\subsubsection{Cross sections}\label{s:numerics:sub4:subsub0}
\begin{figure}[tp]
	\centering
	\includegraphics[width=0.5\textwidth]{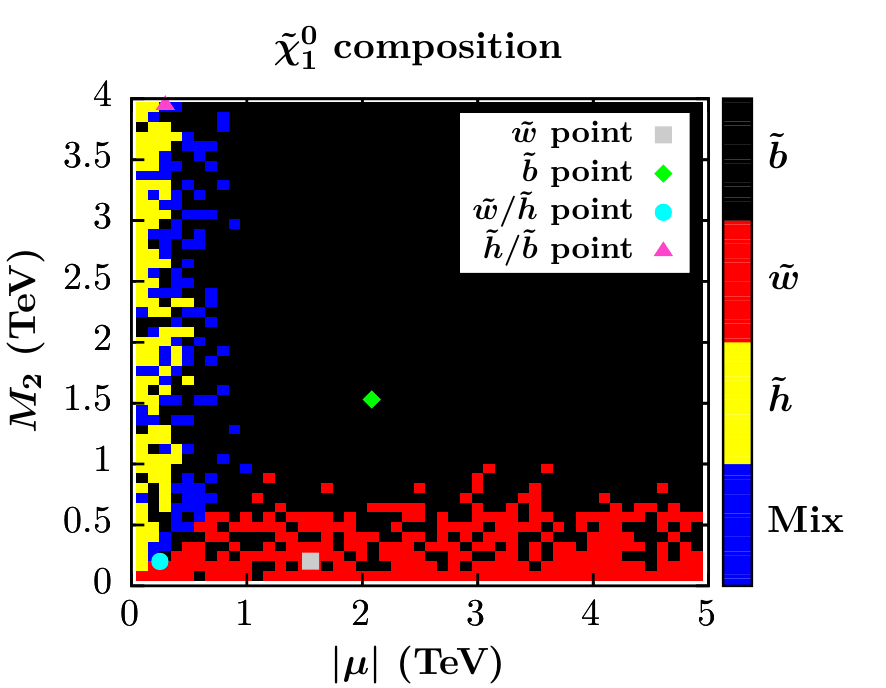}
	\caption{Composition of the lightest neutralino $\tilde{\chi}^0_1$ in the $|\mu|$ -- $M_2$ plane.  A point is dominantly bino-like ($\tilde{b}$, black) for $|N_{11}|^2 > 0.5$, dominantly wino-like ($\tilde{w}$, red) for $|N_{12}|^2 > 0.5$, dominantly higgsino-like ($\tilde{h}$, yellow) for $|N_{13}|^2 > 0.5$ or $|N_{14}|^2 > 0.5$, or a mix of all components (``mix'', blue) in any other case, i.e.\ if neither $|N_{1i}|^2$ is larger than 0.5. 
	For later discussion four specific points referred to as $\tilde{w}, \tilde{b}, \tilde{w}/\tilde{h}$, and $\tilde{h}/\tilde{b}$ are highlighted. 
	}
	\label{pmssmhmmix}
\end{figure}
We move on to discuss the cross section predictions for the case of the more complex pMSSM10 scenario. As we now have a ten-dimensional parameter space, we first must decide which parameters to show on the axes of our ratio plots. We choose to use the modulus of the Higgsino mass parameter $|\mu|$ and the SUSY-breaking wino mass $M_2$, as with this choice it is possible to differentiate between a (predominantly) bino, a wino, and a higgsino composition of $\tilde{\chi}^0_1$. 

In \reffig{pmssmhmmix}, we visualise the magnitude of the mixing matrix elements in the $|\mu|-M_2$ plane with all other pMSSM10 parameters varied implicitly by plotting a point in the shown colour if the corresponding square of the absolute value of the mixing matrix element $|N_{1i}|^2$ is larger than 0.5 (see the caption of the plot for a detailed description of the colouring scheme). Using this colouring scheme, we can see that while the majority of the parameter points in the plane is bino-like (with $|N_{11}|^2 \ge 0.99$), the wino and higgsino components of the lightest neutralino are dominant for low $M_2$ and $|\mu|$, respectively. This is due to the fact that in the lightest neutralino the component with the lightest mass is dominant. Furthermore, we have checked that, with the exception of the higgsino points very close to $|\mu| = 0$~TeV, most of the points labelled ``higgsino'' and ``mix'' up to $|\mu|$ values of around 700~GeV suffer from a higgsino mixing-matrix suppression (c.f.\ discussion at the end of the previous section~\ref{s:numerics:sub3}), leading to comparatively low cross sections in the MSSM and thus to large ratios to the predictions from the simplified models.

\begin{figure}[tp]
	\centering
	\begin{tabular}{cc}
		\includegraphics[width=0.48\textwidth]{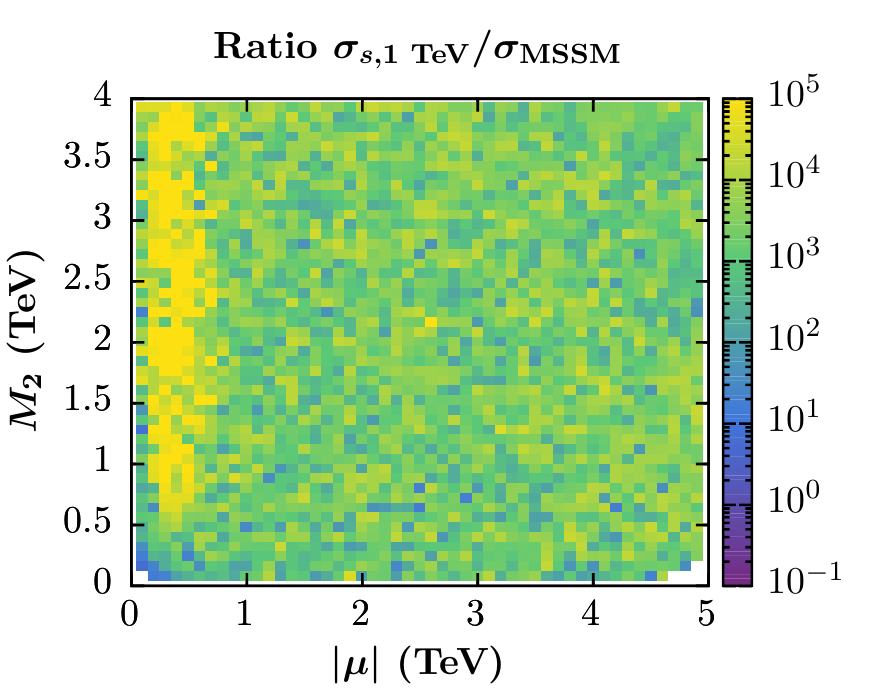} & \includegraphics[width=0.48\textwidth]{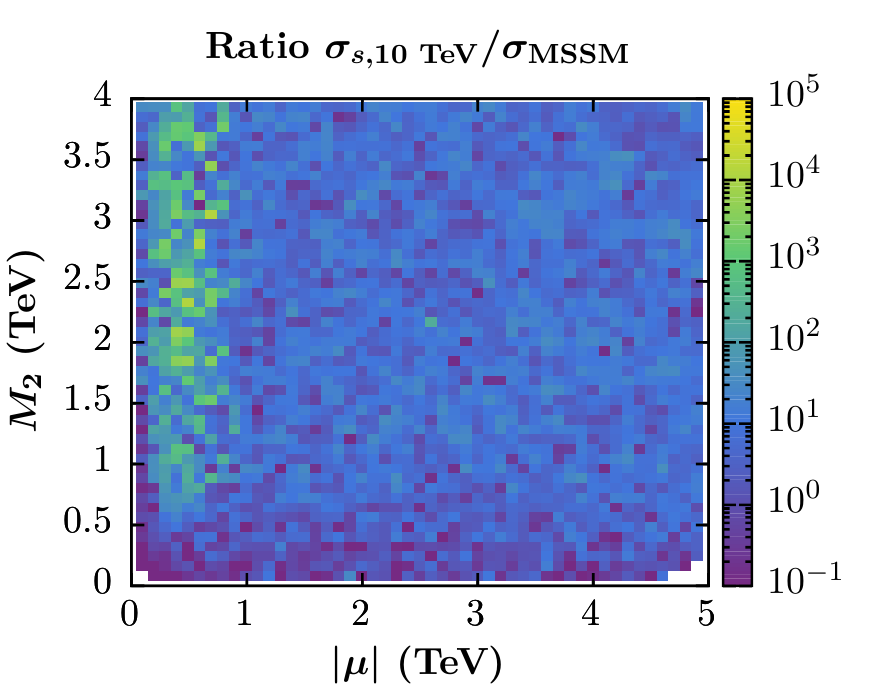}\\
		(a) & (b)\\
		\multicolumn{2}{c}{\includegraphics[width=0.48\textwidth]{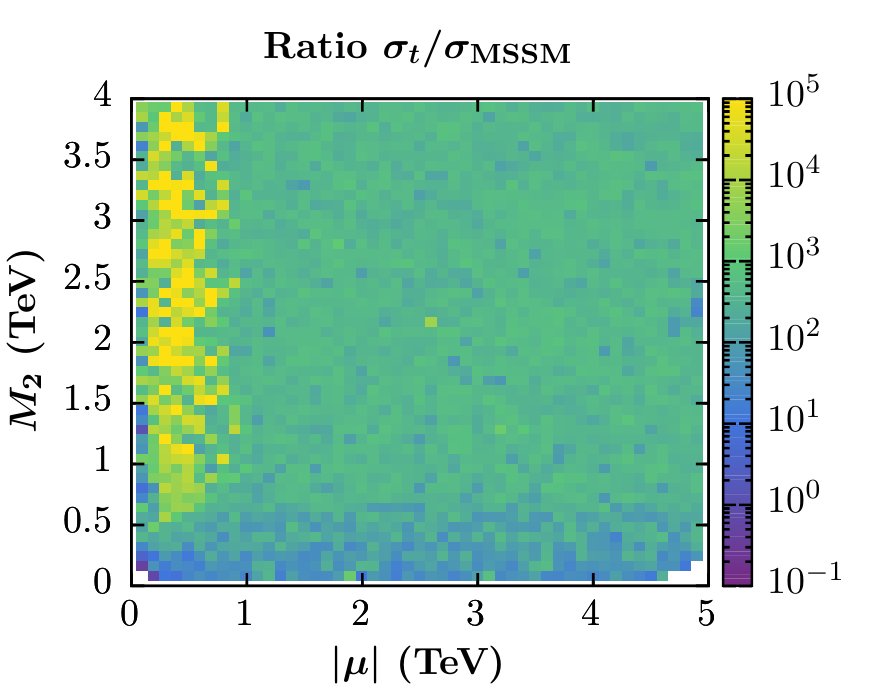}}\\
		\multicolumn{2}{c}{(c)}
	\end{tabular}
	\caption{Ratios of total cross sections for DM pair production in the simplified models to the ones in the MSSM, in the plane of the pMSSM10 parameters $|\mu|$ and $M_2$, with all other parameters varied implicitly. Shown are the $s$-channel models with $M_V = 1$~TeV~(a) and $M_V = 10$~TeV~(b), and the $t$-channel model~(c).}
	\label{pmssmhmplots}
\end{figure}
Similar to the case of the CMSSM,  in \reffig{pmssmhmplots} we present the ratios of total cross sections for DM pair production for simplified model predictions in the numerator, and for MSSM predictions in the denominator, using the axes $|\mu|$ and $M_2$ as motivated in the two previous paragraphs. \Reffigs{pmssmhmplots}~(a) and (b) show the respective ratios for the $s$-channel models with $M_V = 1$~TeV and $M_V = 10$~TeV, respectively. The relative variation of the ratios is very similar for the two different mediator masses, while the absolute values of the ratio for $M_V = 10$~TeV are smaller than for $M_V = 1$~TeV due to the larger suppression from the high mass in the $s$-channel propagator. For $100~\mathrm{GeV} < |\mu| < 700~\mathrm{GeV}$ we observe a region of enhanced ratios which, as discussed before, stems from the higgsino mixing-matrix suppression in the MSSM. For the remaining parameter space, the variation of the ratios appears rather random and varies by two to three orders of magnitude. Both at the leftmost edge of the higgsino region ($|\mu|$ very small) and at the lower edge of the wino region ($M_2$ very small), for the case $M_V = 10$~TeV the ratio becomes smaller than in the bino region, resulting in a larger amount of purple points. This behaviour is related to the value of $m_{\tilde{\chi}^0_1}$ which becomes very small for $|\mu|$ or $M_2$ being close to 0~GeV. For a very low neutralino mass, $m_{\tilde{\chi}^0_1} \lesssim 100$~GeV, the MSSM cross section is enhanced and very sensitive to changes in $m_{\tilde{\chi}^0_1}$ due to the resonance of the $Z$-boson diagram at $m_{\tilde{\chi}^0_1} = M_Z/2$.  For a mediator mass of $M_V = 1$~TeV, the effect is not as pronounced. This is due to the larger sensitivity of the corresponding $s$-channel model itself  to changes in the DM particle mass for low $m_\chi$, because of the lower resonance threshold at $m_\chi = M_V/2 = 500$~GeV. The resonant behaviour therefore is partially smeared out in the ratio with the MSSM cross section.

\Reffig{pmssmhmplots}~(c) shows the ratio of the cross sections in the $t$-channel model to the MSSM cross sections. We can identify similar features as in the case of the $s$-channel model: there is an enhanced region for $100~\mathrm{GeV} < |\mu| < 700~\mathrm{GeV}$ due to the higgsino mixing-matrix suppression in the MSSM, and a suppressed region for low $|\mu|$ or $M_2$ due to the resonant behaviour of the MSSM cross sections (which itself is certainly  absent in the $t$-channel model that does not exhibit resonant topologies like $s$-channel models). However, the bino region is a bit smoother than for the $s$-channel model, varying in most cases by only up to one order of magnitude. This behaviour can again be attributed to the similarity of the $t$-channel model with the MSSM, in particular since a neutralino which is purely bino does not couple to the $Z$-boson and thus the only contributing diagrams are the ones via squark exchange.

In summary, we find that on the basis of the ratios, it is hard to distinguish the $s$- and $t$-channel models from one another, in particular in the regions where the lightest neutralino is predominantly wino- or higgsino-like.

\subsubsection{Differential distributions}\label{s:numerics:sub4:subsub1}
Let us now turn to various differential distributions in the MSSM and the two simplified models considered in this work. As before, for the $s$-channel model we consider mediator masses of  $M_V = 1$~TeV and $M_V = 10$~TeV. While for the $t$-channel model we so far only discussed the case of DM made of Dirac fermions in the previous sections, we now additionally  show results for DM being made of Majorana fermions, which makes the model even more similar to the MSSM. We compare results obtained for these models to those of the MSSM for the four representative points of the pMSSM10 parameter space highlighted in \reffig{pmssmhmmix}.
We focus on the invariant mass of the DM pair system, $M_{\chi\chi}$, and its transverse momentum, $p_{T,\chi\chi}$, which is identical to the ``missing transverse momentum'' in an experimental context.   
Each distribution shown is normalised to the respective total cross section, i.e.\  $\hat{\sigma}_X \equiv (1/\sigma_X)\,d\sigma_X/dy$, where $\hat{\sigma}_X$ is the normalised differential cross section, $\sigma_X$ is the total cross section, $X$ specifies the model, and $y$ denotes either $M_{\chi\chi}$ or $p_{T,\chi\chi}$. For simplicity, no cuts have been applied to any kinematical variable. All results are given at NLO+PS accuracy.

\begin{figure}[tp]
	\centering
	\includegraphics[width=0.495\textwidth]{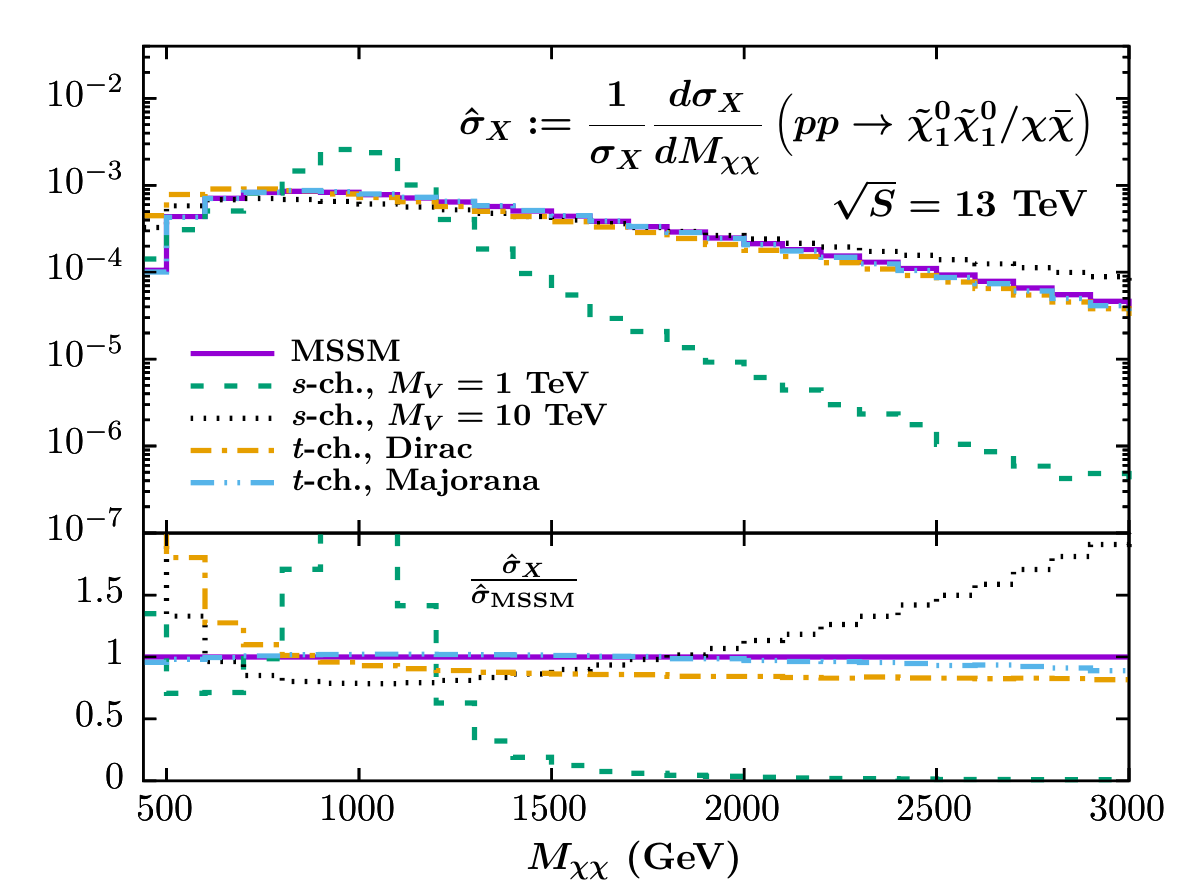}\includegraphics[width=0.495\textwidth]{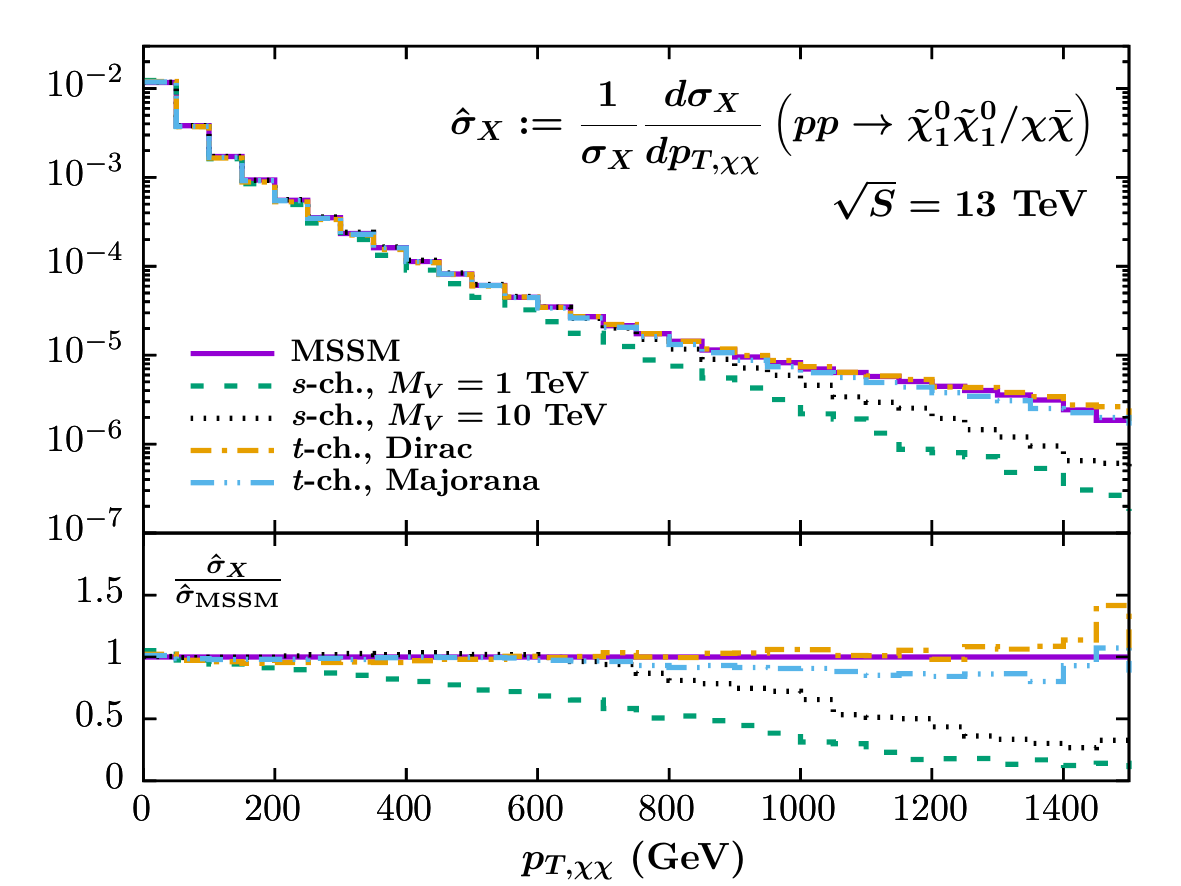}
	\caption{Invariant mass (left) and transverse momentum distribution (right) of the DM pair system at NLO+PS accuracy. Masses and parameters are chosen according to the $\tilde{w}$ parameter point of \reffig{pmssmhmmix}. Shown are predictions for the MSSM (solid purple line), the $s$-channel model with $M_V = 1$~TeV (dashed green line) and $M_V = 10$~TeV (dotted black line), and the $t$-channel model with DM as a Dirac (dash-dotted yellow line) and a Majorana (dash-double dotted blue line) fermion.}
	\label{pmssmwpoint}
\end{figure}
In \reffig{pmssmwpoint}, we present the differential distributions for a parameter point where the lightest neutralino is mainly composed of a wino  (referred to as ``$\tilde{w}$ point'' in \reffig{pmssmhmmix}). We will first focus on the invariant mass distribution of the DM pair system. Even though this quantity is in general not experimentally accessible, as a ``technical variable'' it provides useful information on the structure of the considered scenarios. For the case of the $s$-channel model with $M_V = 1$~TeV we observe a resonance structure peaking at $M_{\chi\chi} = M_V$ which is non-existent for any other model. This effect leads to a strong decline of the distribution towards higher invariant masses, meaning that with the exception of the region around the resonance peak and the lowest bin, the distribution is always lower and dropping more quickly than the one of the MSSM. The case of $M_V = 10$~TeV is a bit different, as there is no resonance structure in the shown $M_{\chi\chi}$ range. For an intermediate $M_{\chi\chi}$ range of 800~GeV~$< M_{\chi\chi} <$ 1300~GeV the distribution approximates the MSSM quite well, as illustrated by the almost constant ratio. For high $M_{\chi\chi}$, however, the distribution falls less rapidly than the MSSM, leading to an increased ratio which strongly deviates from one towards high $M_{\chi\chi}$. The $t$-channel model shows a behaviour which is very close to the MSSM. While for the assumption of Dirac DM, the lowest bins up to $M_{\chi\chi} \approx 1000$~GeV differ from the MSSM, the behaviour for higher invariant masses is essentially the same as in the MSSM, indicated by a constant ratio. A further improvement is seen only for the case of Majorana DM in the $t$-channel model, where the distribution is exactly on top of the prediction of the MSSM.

The right plot of \reffig{pmssmwpoint} shows the transverse-momentum distribution of the DM pair system. We observe that for low $p_{T,\chi\chi} \lesssim 200$~GeV, all models demonstrate the same behaviour. For increasing transverse momentum, the $s$-channel model with $M_V = 1$~TeV is the first to deviate from the MSSM at around $p_{T,\chi\chi} = 200$~GeV with a distribution dropping more quickly than in all of the other scenarios. Beyond $p_{T,\chi\chi} \approx 800$~GeV, also the $s$-channel model with $M_V = 10$~TeV deviates  from the MSSM, as the latter demonstrates a  plateau towards high $p_{T,\chi\chi}$. The plateau originates from the overlap of the decreasing behaviour of the $p_T$ distribution and a resonance peak with its centre at half of the average of the squark masses. The latter is a left-over artifact from the subtraction of on-shell resonances in the real corrections to squark exchange diagrams. Its root is the interference term between the non-resonant and possibly resonant amplitudes $2\operatorname{Re}(\mathcal{M}_{nr}\mathcal{M}_r^*)$ from which no on-shell contributions are subtracted, see section~\ref{s:numerics:sub2:subsub1}.
The predictions of both the $t$-channel models with Dirac DM and Majorana DM, for which the on-shell resonances of the mediators have been subtracted in a similar manner in the code providing the MSSM results, agree well with the $p_T$ distribution of the MSSM in the shown range. The good capability of the $t$-channel model to appproximate the result of the MSSM for this scenario is due to its similarity to the numerically relevant diagrams of the MSSM. As the lightest neutralino is dominantly wino-like, only the squark exchange diagrams contribute, and the $Z$-boson diagram is strongly suppressed.

\begin{figure}[tp]
	\centering
	\includegraphics[width=0.495\textwidth]{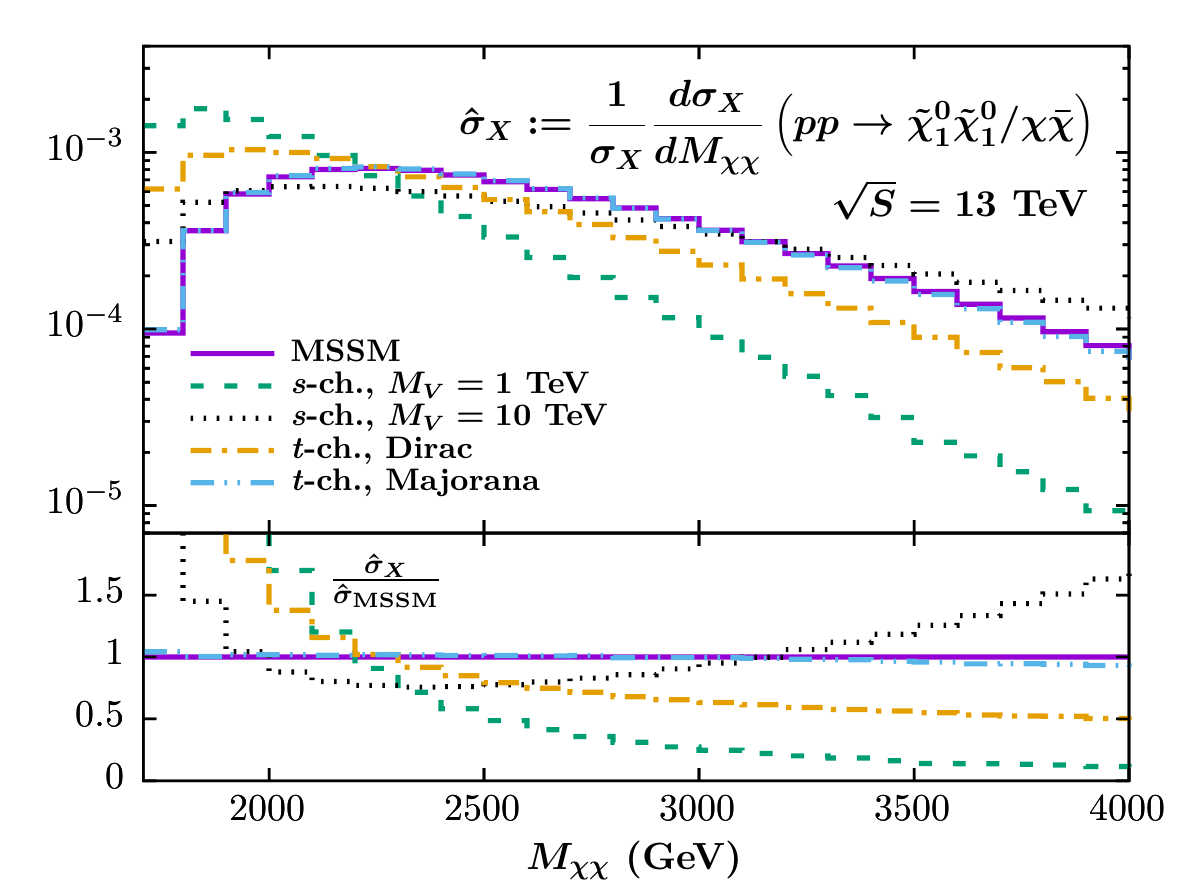}\includegraphics[width=0.495\textwidth]{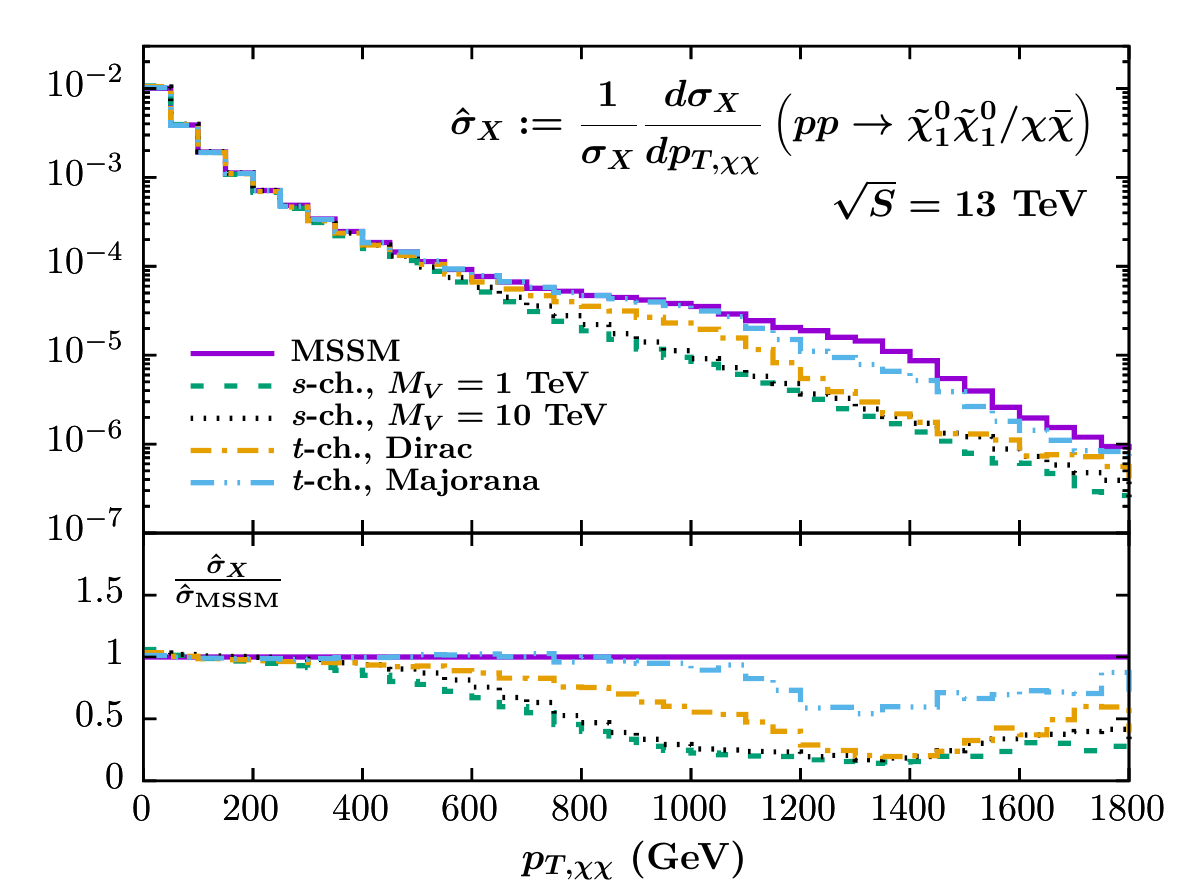}
	\caption{Invariant mass (left) and the transverse momentum distribution (right) of the DM pair system at NLO+PS accuracy. Masses and parameters are chosen according to the $\tilde{b}$ parameter point of \reffig{pmssmhmmix}. The coloring of the lines follows the scheme introduced in \reffig{pmssmwpoint}.}
	\label{pmssmbpoint}
\end{figure}
\Reffig{pmssmbpoint} shows the case of the $\tilde{b}$ point, where the lightest neutralino is dominated by the bino component. The qualitative behaviour of all the lines is very similar to \reffig{pmssmwpoint}. However, while the invariant mass distribution of the $t$-channel model with Majorana DM still lies on top of the MSSM, the case of Dirac DM is now worse, showing a non-constant ratio also for high $M_{\chi\chi}$. The transverse momentum distributions of the $s$-channel models with two different mediator masses are now indistinguishable, which is related to a higher DM mass than for the parameter point considered in \reffig{pmssmwpoint}. We furthermore observe that now the $t$-channel model with Dirac DM does not describe the shape of the MSSM perfectly anymore.

\begin{figure}[tp]
	\centering
	\includegraphics[width=0.495\textwidth]{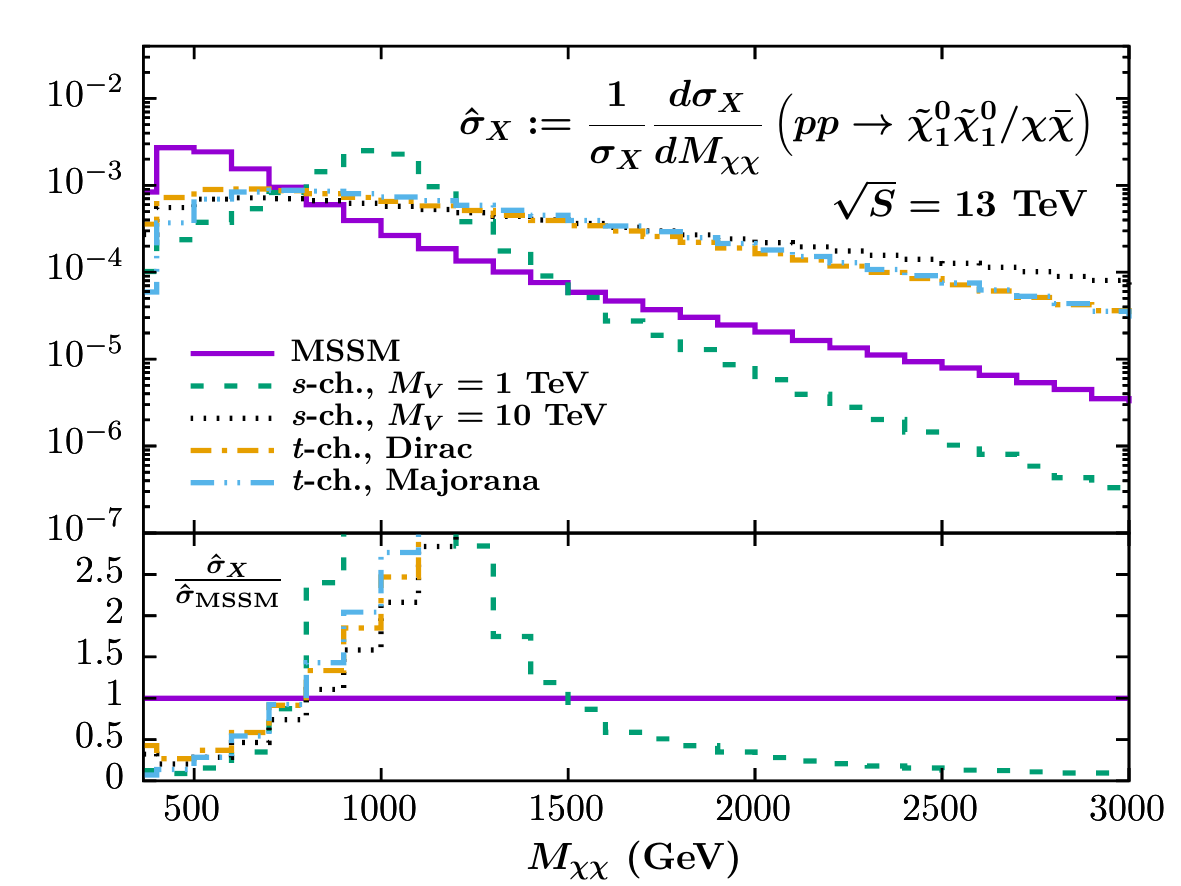}\includegraphics[width=0.495\textwidth]{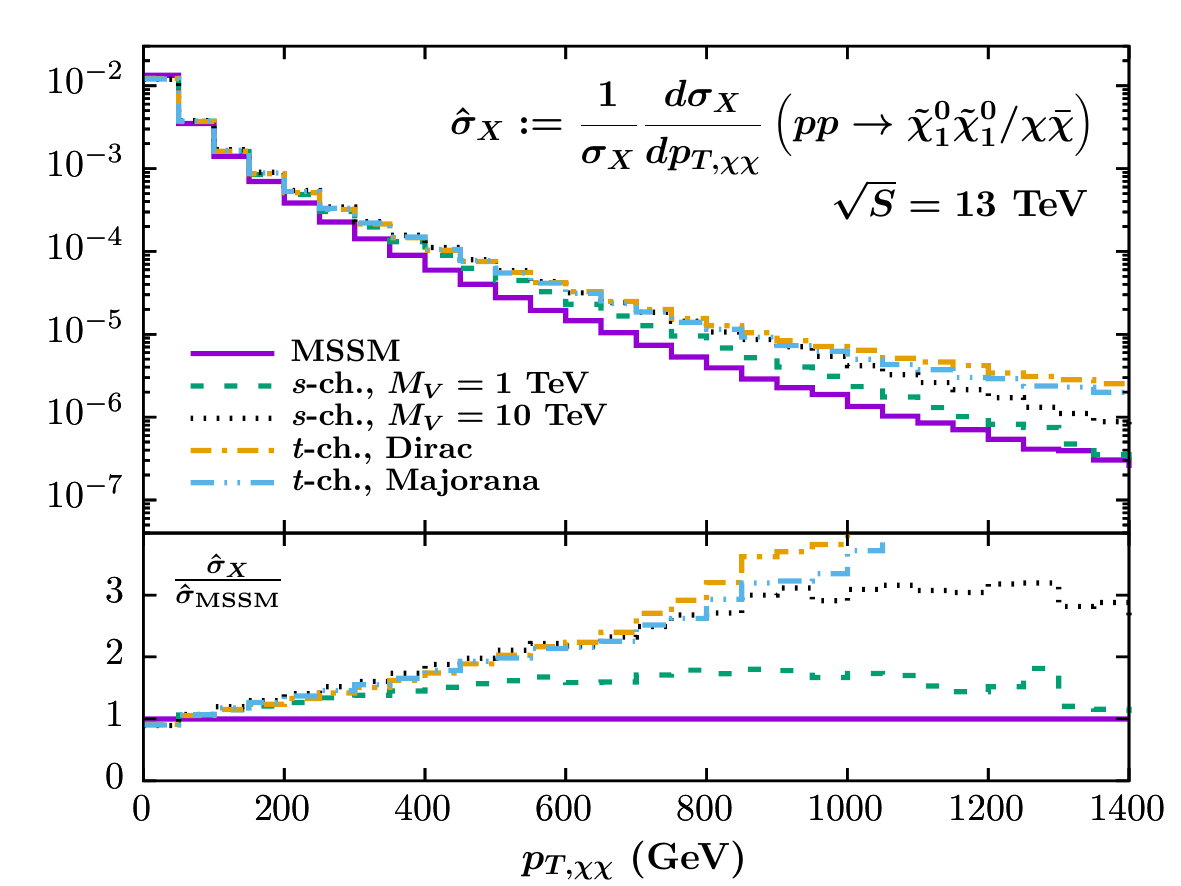}
	\caption{Invariant mass (left) and the transverse momentum distribution (right) of the DM-pair system at NLO+PS accuracy. The masses and parameters are chosen according to the $\tilde{w}/\tilde{h}$ parameter point of \reffig{pmssmhmmix}. The coloring of the lines follows the scheme introduced in \reffig{pmssmwpoint}.}
	\label{pmssmwhpoint}
\end{figure}
Differential distributions for the $\tilde{w}/\tilde{h}$ point are shown in \reffig{pmssmwhpoint}. Here, the lightest neutralino consists of mix between a dominant wino component and a small higgsino part. The invariant mass distribution illustrates that none of the simplified models is able to reproduce the behaviour of the MSSM: while the $s$-channel model with $M_V = 1$~TeV fails to describe the MSSM line due to the resonance behaviour of the mediator, the other three lines are either below the MSSM prediction for low invariant masses or far above towards high invariant masses. A similar observation can be made for the $p_{T,\chi\chi}$ distribution: While all the lines of the simplified models decrease more slowly than the one of the MSSM, the ratio for the $s$-channel models is constant towards high $p_{T,\chi\chi}$. In particular the ratio of the $s$-channel model with $M_V = 1$~TeV appears to be approximately constant for 300~GeV $< p_{T,\chi\chi} <$ 1200~GeV.
The previous results indicate that even for a parameter point where the wino component is dominant ($|N_{12}|^2 > 0.6$), a small, but non-vanishing and non-cancelling higgsino component can lead to a strong influence of the $Z$-boson diagram, which manifests itself in a poor capability of the $t$-channel model, but (at least for the $p_{T,\chi\chi}$ distribution) a relatively good capability of one of the $s$-channel models to approximate the behaviour of the MSSM. 

\begin{figure}[tp]
	\centering
	\includegraphics[width=0.495\textwidth]{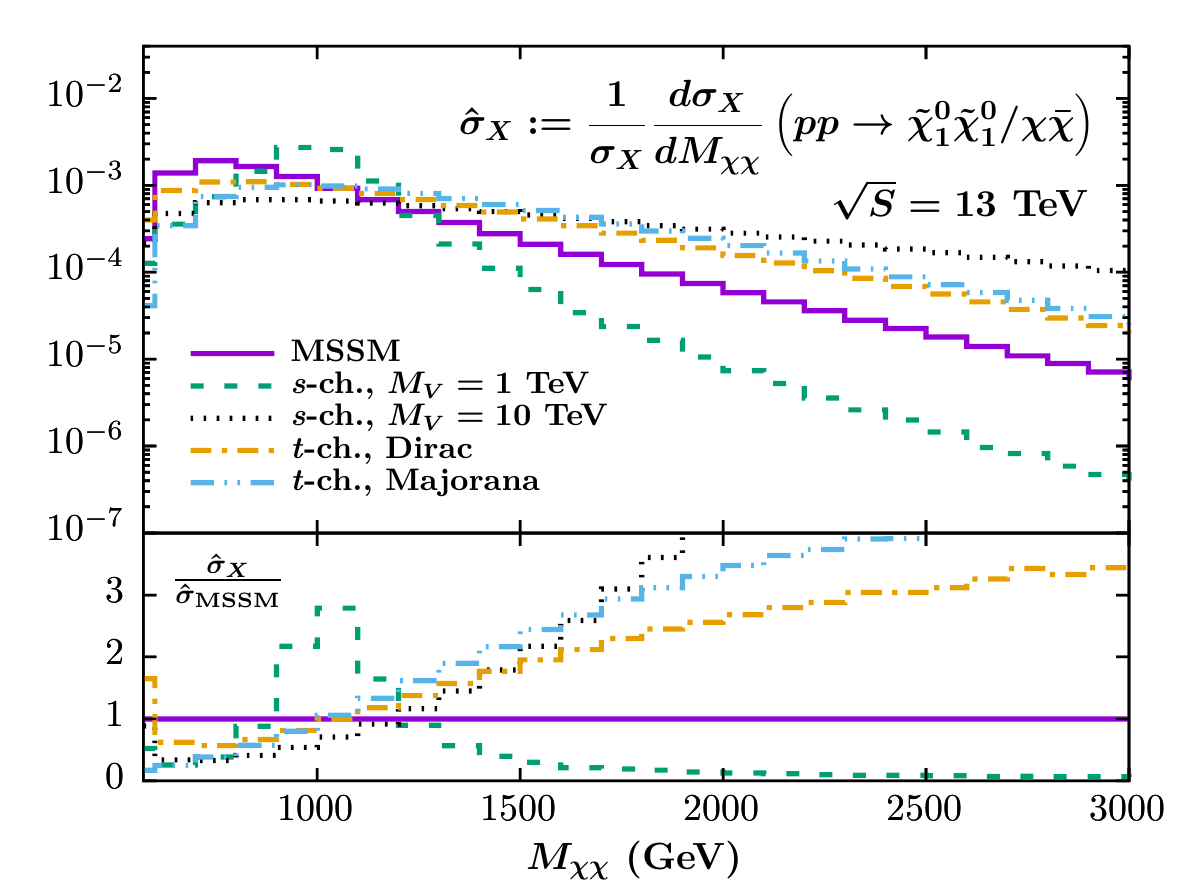}\includegraphics[width=0.495\textwidth]{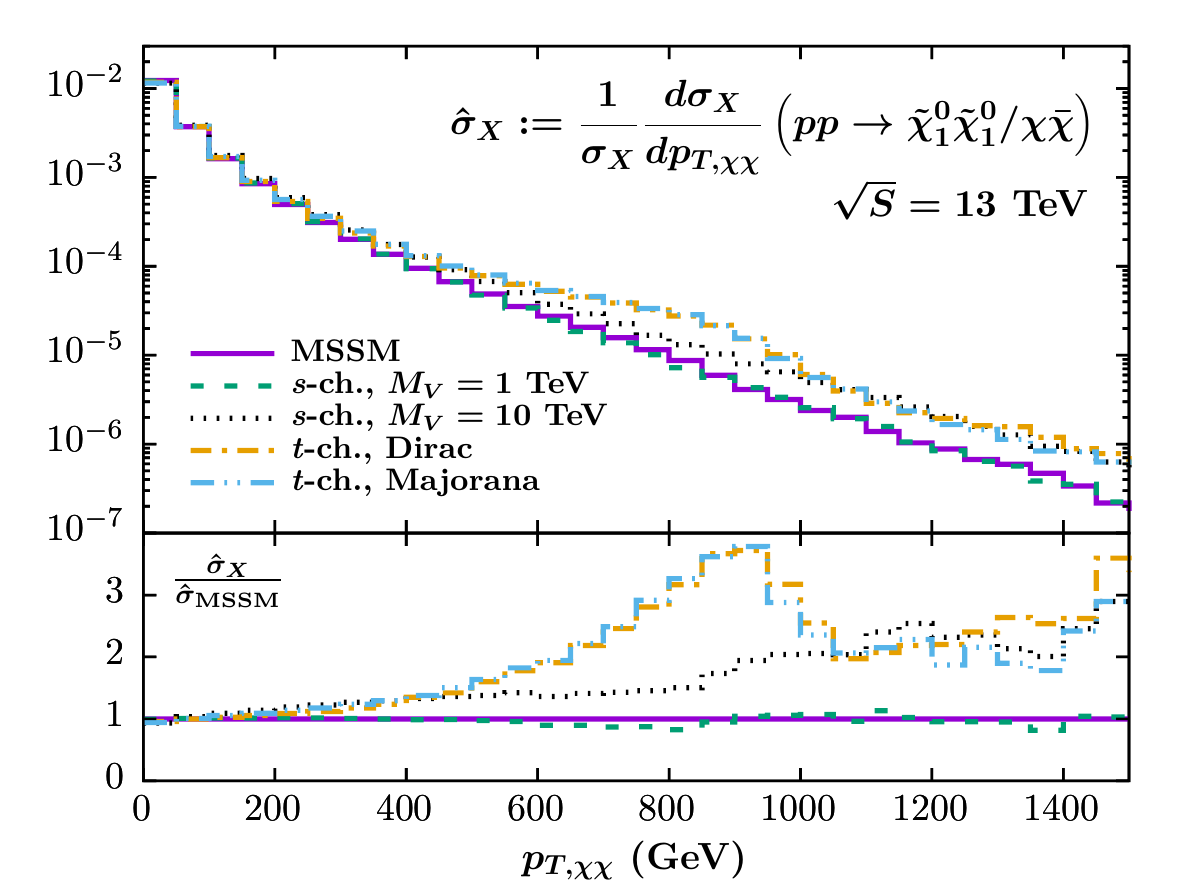}
	\caption{Invariant mass (left) and the transverse momentum distribution (right) of the DM pair system at NLO+PS accuracy. Masses and parameters are chosen according to the $\tilde{h}/\tilde{b}$ parameter point of \reffig{pmssmhmmix}. The coloring of the lines follows the scheme introduced in \reffig{pmssmwpoint}.}
	\label{pmssmhbpoint}
\end{figure}
Finally, predictions for the $\tilde{h}/\tilde{b}$ point are shown in \reffig{pmssmhbpoint}. For this point, the lightest neutrino is mainly a mix of the bino ($|N_{11}|^2 \approx 0.13$) and higgsino ($|N_{13}|^2 \approx 0.45$, $|N_{14}|^2 \approx 0.41$) components, while the wino component is negligible. The invariant mass distribution reveals an equally bad behaviour of all the simplified model predictions to account for the MSSM like in the previous parameter point. Interestingly, the $p_{T,\chi\chi}$ distribution of the $s$-channel model with $M_V = 1$~TeV follows the MSSM prediction almost perfectly, while all of the other lines lie above the MSSM with a non-constant ratio. In particular, the $t$-channel models show a peak-like structure at around $p_{T,\chi\chi} \approx 900$~GeV, which is related to the on-shell subtraction of intermediate $t$-channel mediators, as discussed previously. For a parameter point where the MSSM prediction is dominated by the $Z$-boson exchange diagram, such a structure is not visible, as the squark exchange diagrams are suppressed.
%

\section{Conclusions and outlook}\label{s:conclusion}
In this paper we studied in detail two simplified models for DM. In these models, our DM candidate is a fermion, interacting with the SM via a vector which is a singlet under the SM gauge group ($s$-channel model) or via a flavoured scalar ($t$-channel model). In the first case we considered a Dirac fermion, while in the latter we also built a model with a Majorana fermion, in order to compare with supersymmetric theories, where the DM candidate is a Majorana fermion, the lightest neutralino.

We implemented the hadronic DM pair production processes within these simplified models in the \POWHEGBOX{} framework at NLO+PS level. Using this implementation, we compared the production cross sections and distributions of these models with the ones of neutralino pair-production in the MSSM. Our analysis shows that the simplified models are capable of reconstructing the supersymmetric case to some extent but never in all considered kinematic regions and for arbitrary points of the MSSM parameter space. 

For the parameter point where the higgsino component of the lightest neutralino is negligible the $t$-channel simplified model produces results that are rather similar to the ones predicted by the MSSM (c.f.\ \reffigs{pmssmwpoint} and \ref{pmssmbpoint}). Using a Majorana fermion as a DM candidate leads to an additional improvement of the approximation. This does not come as a surprise, since the lack of a higgsino component suppresses $s$-channel contributions within the MSSM, therefore making the numerically relevant topologies of the $t$-channel simplified model and of the MSSM the same. 

In contrast, the presence of a higgsino component worsens considerably the agreement between the simplified  models  and the MSSM (c.f.\ \reffigs{pmssmwhpoint} and \ref{pmssmhbpoint}). In the MSSM, $s$-channel contributions due to this component have a large impact, because the process is favoured by the relatively low mass of the $Z$-boson that mediates it. In the simplified $s$-channel model, the mediator is much more massive. Therefore this channel has a rather different behaviour, leading to the unsatisfactory agreement between this model and the MSSM that we showed. 

In summary, it seems difficult to reproduce the complete phenomenology of neutralino pair production for DM studies with a simplified model with such a restricted set of parameters like the ones we showed in this work. The neutralino mixing matrix of the MSSM is crucial in its phenomenology, but it does not have an equivalent in the simple models we considered. Consequently, the reduction of the MSSM to a simpler model is not a trivial issue and needs to be addressed carefully, either recurring to a higher degree of complexity in the model building or considering only specific MSSM scenarios.

Nevertheless, we have checked that the event rates associated with the $p_T$ distributions of section~\ref{s:numerics:sub4:subsub1} would be too small for the differences between the different models to be detectable. In these regions of $p_T$, no events are expected to be observed at the LHC or at the planned HL-LHC upgrade, making the models indistinguishable from an experimental point of view. Simplified models are then a usable and efficient tool in DM searches at colliders at the current stage. The $p_T$ regions with a sizeable disagreement between the models could, however, be probed by a future high-energy collider with a centre-of-mass energy of 100 TeV.

\section*{Acknowledgements}
The authors would like to thank Julien Baglio, Matthias Kesenheimer, and Junya Nakamura for valuable discussions. Part of this work was performed on the high-performance computing resource bwForCluster NEMO with support by the state of Baden-W\"urttemberg through bwHPC and the German Research Foundation (DFG) through grant no INST 39/963-1 FUGG. The authors acknowledge support by the Institutional Strategy of the University of T\"ubingen (DFG, ZUK 63), and the Carl Zeiss Foundation.

\providecommand{\href}[2]{#2}\begingroup\raggedright\endgroup

\end{document}